\documentclass[pra,aps,10pt]{revtex4-2}
\usepackage{makeidx}
\usepackage{amsmath,amssymb,graphicx}
\usepackage{color}
\usepackage{subfigure}

\begin{document}

\title{Trapping wave fields in an expulsive potential by means of linear
coupling}
\author{Nir Hacker$^1$ and Boris A. Malomed$^{1,2}$}

\affiliation{$^1$Department of Physical Electronics, School of Electrical Engineering, 
Faculty of Engineering, and Center for Light-Matter interaction, Tel Aviv University,
Tel Aviv 69978, Israel}
\affiliation{$^2$Instituto de Alta
Investigaci\'{o}n, Universidad de Tarapac\'{a}, Casilla 7D, Arica, Chile}

\begin{abstract}
We demonstrate the existence of confined states in one- and two-dimensional
(1D and 2D) systems of two linearly-coupled components, with the confining
harmonic-oscillator (HO) potential acting upon one component, and an
expulsive anti-HO potential acting upon the other. The systems can be
implemented in optical and BEC dual-core waveguides. In the 1D linear
system, codimension-one solutions are found in an exact form for the ground
state (GS) and dipole mode (the first excited state). Generic solutions are
produced by means of the variational approximation, and are found in a
numerical form. Exact codimension-one solutions and generic numerical ones
are also obtained for the GS and vortex states in the 2D system (the exact
solutions are found for all values of the vorticity). Both the trapped and
anti-trapped components of the bound states may be dominant ones, in terms
of the norm. The localized modes may be categorized as \textit{bound states
in continuum}, as they coexist with delocalized ones. The 1D states, as well
as the GS in 2D, are weakly affected and remain stable if the
self-attractive or repulsive nonlinearity is added to the system. The
self-attraction makes the vortex states unstable against splitting, while
they remain stable under the action of the self-repulsion.
\end{abstract}

\maketitle

\section{Introduction}

The combination of the harmonic-oscillator (HO) trapping potential and cubic
nonlinearity is a ubiquitous setting which occurs in diverse physical
realizations. A well-known one is offered by Bose-Einstein condensates
(BECs) with the mean-field nonlinearity \cite{Pet,Pit,Pan}, loaded in a
magnetic or optical trap -- see, e.g., Refs.~\cite{Grossmann}-\cite{PLA}. A
similar combination of the effective confinement, approximated by the
parabolic profile of the local refractive index, and the Kerr term is
relevant as a model of optical waveguides \cite{Agrawal}-\cite{Thaw}. Models
of the same type find other physical realizations too, such as polariton
condensates \cite{polaritons} and networks of Josephson oscillators \cite%
{Josephson network}.

The interplay of the self-attractive nonlinearity (self-focusing, in terms
of optics) and trapping potential gives rise to localized modes, which may
be considered as bright solitons confined by the external potential \cite%
{KA,Peyrard}. On the other hand, dynamics of solitons in expulsive
potentials, such as an inverted HO (\textit{anti-HO}), is also relevant in
various physical settings \cite{Carr}-\cite{KartKon}. In terms of optics,
expulsive potentials represents anti-waveguiding setups, which are used in
the design of photonic data-processing schemes \cite{anti-WG1}-\cite{Kaplan}%
. It is worthy to mention that the interplay of the self-repulsive
nonlinearity and a combination of spatially periodic (lattice) and HO or
anti-HO potentials gives rise to gap solitons with a negative effective
mass, which are expelled by the normal HO potential, and stay trapped under
the action of the anti-HO one \cite{HS}.

The objective of this work is to introduce a system of linearly coupled one-
and two-dimensional (1D and 2D) \textquotedblleft cores" (conduits for
photonic or matter waves), with \textit{opposite signs} of HO potentials
acting in them, one confining and one expulsive. In BEC, such a system can
be built as a pair of parallel cigar- or pancake-shaped condensates coupled
by tunneling of atoms, which may be enhanced by an additional optical link
\cite{Ketterle}. Then, the HO and anti-HO potentials in the cores can be
induced by properly shaped red- and blue-detuned laser beams illuminating
the cores (see. e.g., Ref. \cite{two-color}). In optics, an effectively 1D
coupler composed of planar waveguiding and antiwaveguiding cores can be
fabricated in a straightforward way \cite{polymers}, although a 2D version
of such an optical system is not a realistic one.

We consider these systems both in the linear regime and with inclusion of
intra-core self-attractive or repulsive cubic nonlinearity. In fact, the
linear version of the system is the most relevant one, as the existence of a
trapped states in the system with linearly-coupled components subject to the
action of the confining and expulsive potentials is not obvious -- in
particular, because the straightforward definition of spectra does not
directly apply to such a system. A \textit{half-trapped }linearly-coupled
system, including the trapping HO potential in one component and no
potential in the other, was considered recently \cite{we}, but, to the best
of our knowledge, trapped-antitrapped systems were not addressed before.

It is obvious that, irrespective of the existence of confined states in the
system with the anti-trapping component, it also gives rise to a continuous
family of delocalized states, therefore the confined ones, if they exist,
may be considered as a specific case of \emph{bound states in the continuum}
(BIC)\ \cite{BIC,BIC2,SOC}, alias \emph{embedded modes} \cite{embedded}.
Recently, applications of BIC of various types have drawn much interest in
photonics \cite{BIC3}-\cite{BIC5}.

The presentation is organized below as follows. The 1D and 2D systems are
introduced in Section II. Then, section III reports exact analytical
solutions of the \textit{codimension-one} type (with one constraint imposed
on the system's parameters), for both 1D and 2D systems in their linear
form. These are ground-state (GS) and dipole-mode (DM) solutions in 1D, and
2D states with all integer values of vorticity, $S\geq 0$ ($S=0$ represents
the GS in 2D, while $S=1$ is the 2D counterpart of the DM). The exact
solutions play an important role, as they provide an \textit{exact proof} of
the existence of the bound state in the presence of the expulsive potential
in one component. Generic, although approximate, analytical results and
their numerical counterparts are produced, for the linear systems, in
Section IV. The analytical results provide eigenvalues of the propagation
constant, in terms of optics (or chemical potential, in BEC) for the GS and
DM states in 1D, predicted by the variational (Rayleigh-Ritz) approximation.
Another analytical finding is an asymptotic form, at $|x|\rightarrow \infty $
(or $r\rightarrow \infty $) of the 1D (or 2D) delocalized states. Effects of
the nonlinearity are considered, in the numerical form, in Section V, which
addresses stationary GS, DM, and 2D vortex modes in the nonlinear systems,
and simulations of the evolution of inputs produced by exact eigenmodes of
the linear system under the action of the nonlinearity. In the latter case,
the system demonstrates the emergence of robust breathers for moderate
nonlinearity, and chaotization when it is too strong. The vortex states in
2D are, as usual \cite{PhysD}, unstable against spontaneous splitting under
the action of self-attraction, and remain stable in the case of
self-repulsion. The paper is concluded by Section VI.

\section{The one- and two-dimensional systems}

The 1D\ system is introduced, in the scaled form, as a pair of
linearly-coupled Schr\"{o}dinger equations with the HO and anti-HO
potentials in the first and second components. The equations are written in
the notation adopted in optics, with propagation distance $z$ and transverse
coordinate $x$:

\begin{eqnarray}
iu_{z}+\frac{1}{2}u_{xx}+\lambda v-\frac{1}{2}x^{2}u+\sigma |u|^{2}u
&=&-\omega u,  \label{u} \\
iv_{z}+\frac{1}{2}v_{xx}+\lambda u+\frac{1}{2}\kappa x^{2}v+\sigma |v|^{2}v
&=&0.  \label{v}
\end{eqnarray}%
Here the diffraction coefficients and strength of the HO\ trapping potential
acting on the $u$ component are normalized to be $1$, $\lambda $ is the
coefficient of the linear coupling, which may be set to be positive, $\kappa
>0$ is the strength of the expulsive potential acting on $v$, and $\omega $
is a possible propagation-constant mismatch between the components. Finally,
$\sigma =\pm 1$ or $0$ represents the scaled nonlinearity coefficient, $%
\sigma =+1$ and $-1$ corresponding to the self-focusing and defocusing signs
of the nonlinearity.

Stationary solutions to Eqs.~(\ref{u}) and (\ref{v}) are looked for as%
\begin{equation}
\left\{ u,v\right\} =\left\{ U(x),V(x)\right\} \exp \left( -i\mu z\right) ,
\label{uvUV}
\end{equation}%
with real propagation constant $-\mu $ (in BEC, with $z$ replaced by scaled
time $t$, $\mu $ is the chemical potential), and real functions $U(x)$ and $%
V(x)$ satisfying equations%
\begin{gather}
\left( \mu +\omega \right) U+\frac{1}{2}\frac{d^{2}U}{dx^{2}}+\lambda V-%
\frac{1}{2}x^{2}U+\sigma U^{3}=0,  \label{Ufull} \\
\mu V+\frac{1}{2}\frac{d^{2}V}{dx^{2}}+\lambda U+\frac{1}{2}\kappa
x^{2}V+\sigma V^{3}=0.  \label{Vfull}
\end{gather}

In fact, the nonlinearity plays a secondary role in the present context, the
most important issue being the existence of trapped modes in the linear
version of Eqs. (\ref{Ufull}) and (\ref{Vfull}), which corresponds to $%
\sigma =0$. The main objective of the analysis presented below is finding
spectra of eigenvalues $\mu $ for such modes.

The 2D version of the system, written in the\ polar coordinates $\left(
r,\theta \right) $, is%
\begin{eqnarray}
iu_{z}+\frac{1}{2}\left( \frac{\partial ^{2}}{\partial r^{2}}+\frac{1}{r}%
\frac{\partial }{\partial r}-\frac{1}{r^{2}}\frac{\partial ^{2}}{\partial
\theta ^{2}}\right) u+\lambda v-\frac{1}{2}r^{2}u+\sigma |u|^{2}u &=&-\omega
u,  \label{u2D} \\
iv_{z}+\frac{1}{2}\left( \frac{\partial ^{2}}{\partial r^{2}}+\frac{1}{r}%
\frac{\partial }{\partial r}-\frac{1}{r^{2}}\frac{\partial ^{2}}{\partial
\theta ^{2}}\right) v+\lambda u+\frac{1}{2}\kappa r^{2}v+\sigma |v|^{2}v
&=&0.  \label{v2D}
\end{eqnarray}%
Stationary solutions to Eqs. (\ref{u2D}) and (\ref{v2D}) with integer
vorticity $S$ are looked for as%
\begin{equation}
\left\{ u,v\right\} =\exp \left( -i\mu z+iS\theta \right) \left\{
U(r),V(r)\right\} ,  \label{vortex}
\end{equation}%
where real functions $U$ and $V$ satisfy radial equations%
\begin{gather}
\left( \mu +\omega \right) U+\frac{1}{2}\left( \frac{d^{2}}{dr^{2}}+\frac{1}{%
r}\frac{d}{dr}-\frac{S^{2}}{r^{2}}\right) U+\lambda V-\frac{1}{2}%
r^{2}U+\sigma U^{3}=0,  \label{U2D} \\
\mu V+\frac{1}{2}\left( \frac{d^{2}}{dr^{2}}+\frac{1}{r}\frac{d}{dr}-\frac{%
S^{2}}{r^{2}}\right) V+\lambda U+\frac{1}{2}\kappa r^{2}V+\sigma V^{3}=0.
\label{V2D}
\end{gather}

Conserved quantities of the 1D system are the integral power (norm) and
Hamiltonian
\begin{gather}
P=\int_{-\infty }^{+\infty }\left[ \left\vert u(x)\right\vert
^{2}+\left\vert v(x)\right\vert ^{2}\right] dx\equiv P_{u}+P_{v},  \label{P}
\\
H=\int_{-\infty }^{+\infty }\left[ \frac{1}{2}\left( \left\vert
u_{x}\right\vert ^{2}+\left\vert v_{x}\right\vert ^{2}+|u|^{2}-\kappa
|v|^{2}\right) -\lambda \left( u^{\ast }v+uv^{\ast }\right) +\frac{\sigma }{2%
}\left( |u|^{4}+|v|^{4}\right) \right] dx.  \label{H}
\end{gather}%
The 2D system conserves obvious counterparts of integral quantities (\ref{P}%
) and (\ref{H}), and also the angular momentum,%
\begin{equation}
M=-i\int_{0}^{\infty }rdr\int_{0}^{2\pi }d\theta \left( u^{\ast }\frac{%
\partial u}{\partial \theta }+v^{\ast }\frac{\partial v}{\partial \theta }%
\right) ,  \label{M}
\end{equation}%
where $\ast $ stands for the complex conjugate. Obviously, for eigenstates (%
\ref{vortex}) Eq. (\ref{M}) have $M=SP$.

The evolution of the wave fields was simulated, in the framework of Eqs. (%
\ref{u}), (\ref{v}) and (\ref{u2D}), (\ref{v2D}), with the help of the
MATLAB's ode15s program, which is a variable-step, variable-order solver
based on the numerical differentiation formulas (NDFs) of orders up to $5$.
The spatial discretization was performed with a variable mesh size $\Delta
x=\Delta y\lesssim 0.01$, and the propagation distance was discretized with
a variable step size $\Delta z\lesssim 0.1$. Stationary equations (\ref%
{Ufull}), (\ref{Vfull}) and (\ref{U2D}), (\ref{V2D}) were solved numerically
by means of the collocation method, based on the Runge-Kutta algorithm, with
the same values of $\Delta x$ and $\Delta y$. All the equations were solved
with the zero (Dirichlet) boundary conditions.

The size of $\Delta x$, $\Delta y$, and $\Delta z$, as well as the NDF
order, were dynamically adjusted at each iteration step of the simulations,
according to local error estimation, which had to satisfy the predefined
tolerance. In the present work, the relative tolerance (pertaining to the
local solution value) and the absolute tolerance were fixed as $10^{-3}$ and
$10^{-6}$, respectively. This choice made it possible to secure the
stability of the numerical scheme. In particular, no visible change in the
solutions was observed if the calculations were rerun with smaller values of
$\Delta x$, $\Delta y$ (up to $0.001$), and $\Delta z$. In fact, some
numerical results are plotted with values of $\Delta x$, $\Delta y$, and $%
\Delta z$ somewhat larger than those which were used in the simulations, as
very small values were not necessary for plotting.

\section{Exact solutions for bound states\textit{\ }in the linear system}

\subsection{The ground state (GS)}

The linear version of the 1D stationary system, represented by Eqs. (\ref%
{Ufull}) and (\ref{Vfull}) with $\sigma =0$, admits a \textit{codimension-one%
} (non-generic) exact solution for the system's GS:

\begin{gather}
U(x)=\left( U_{0}+U_{2}x^{2}\right) \exp \left( -\frac{x^{2}}{2}\right) ,
\label{U(x)} \\
V(x)=V_{0}\exp \left( -\frac{x^{2}}{2}\right) ,  \label{V(x)} \\
U_{0}=\frac{1+\kappa -2\lambda ^{2}}{4\lambda }V_{0},  \label{U0} \\
U_{2}=-\frac{1+\kappa }{2\lambda }V_{0},  \label{U2} \\
\mu _{\mathrm{GS}}=\frac{1}{2}\left( \lambda ^{2}+\frac{1}{2}\right) -\frac{%
\kappa }{4},  \label{mu}
\end{gather}%
where $V_{0}$ is an arbitrary amplitude. The codimension-one character of
this solution is determined by the fact that it exists only if the following
relation is imposed on parameters $\omega ,$ $\kappa $, and $\lambda $:
\begin{equation}
\omega _{\mathrm{GS}}=\frac{9}{4}-\frac{\lambda ^{2}}{2}+\frac{\kappa }{4}
\label{omega}
\end{equation}%
(cf. Refs. \cite{embedded} and \cite{Avila}, as concerns the definition of
the codimensionality one). Note that, as it follows from Eqs. (\ref{U(x)}%
), (\ref{U0}), and (\ref{U2}), component $U(x)$ of the exact solution
crosses zero at $x^{2}=\left( 1+\kappa -2\lambda ^{2}\right) /[2(1+\kappa )]$%
, under the condition of
\begin{equation}
\lambda ^{2}<(1/2)(1+\kappa ),  \label{0-crossing}
\end{equation}%
Further, in the interval of
\begin{equation}
(1/2)(1+\kappa )<\lambda ^{2}<(5/2)(1+\kappa ),  \label{<<}
\end{equation}%
$U(x)$ does not change its sign, but exhibits a local minimum at $x=0$. An example of
this feature is shown below, in the right panel of Fig. \ref{fig1}.

A ``naive assumption" may be that the confined mode would be
maintained by the linear coupling between $U$ and $V$, in spite of the
action of the expulsive potential upon the $V(x)$ component, if the
naturally trapped one, $U(x)$, dominates in the system, i.e., if its
integral power exceeds the power of $V$,
\begin{equation}
P_{u}/P_{v}>1,  \label{>1}
\end{equation}%
see Eq. (\ref{P}). Nevertheless, the confined mode may exist even if
condition (\ref{>1}) does not hold. Indeed, for the exact solution (\ref%
{U(x)})-(\ref{U2}) the power ratio is%
\begin{equation}
\left( \frac{P_{u}}{P_{v}}\right) _{\mathrm{GS}}=\frac{\lambda ^{2}}{4}+%
\frac{(1+\kappa )^{2}}{8\lambda ^{2}}.  \label{P/P}
\end{equation}%
Straightforward analysis demonstrates that Eq. (\ref{P/P}) always yields $%
P_{u}/P_{v}>1$ if the strength of the expulsive potential is large enough,
\textit{viz}.,
\begin{equation}
\kappa >2\sqrt{2}-1\approx 1.83,  \label{kappa>}
\end{equation}%
which agrees with the above-mentioned \textquotedblleft naive" expectation.
On the other hand, if condition (\ref{kappa>}) does not hold, Eq. (\ref{P/P}%
) gives $P_{u}/P_{v}<1$ in the following interval of values of the coupling
constant:%
\begin{equation}
2-\sqrt{4-(1+\kappa )^{2}/2}<\lambda ^{2}<2+\sqrt{4-(1+\kappa )^{2}/2}.
\label{<lambda<}
\end{equation}

\subsection{The dipole mode (first excited state)}

In addition to the spatially-even GS solution, an exact odd one (DM), which
represents the first excited state of the system, can be found too:%
\begin{gather}
U(x)=\left( U_{1}x+U_{3}x^{3}\right) \exp \left( -\frac{x^{2}}{2}\right) ,
\label{Uodd} \\
V(x)=V_{1}x\exp \left( -\frac{x^{2}}{2}\right) ,  \label{Vodd} \\
U_{1}=\frac{3(1+\kappa )-2\lambda ^{2}}{4\lambda }V_{1},  \label{U1} \\
U_{3}=-\frac{1+\kappa }{2\lambda }V_{1},  \label{U3} \\
\mu _{\mathrm{DM}}=\frac{1}{2}\left( \lambda ^{2}+\frac{3}{2}\right) -\frac{%
3\kappa }{4},  \label{mu-odd}
\end{gather}%
where $V_{1}$ is an arbitrary amplitude. It is also a codimension-one
solution, which exists if the following constraint is imposed onto the
parameters:%
\begin{equation}
\omega _{\mathrm{DM}}=\frac{11}{4}-\frac{\lambda ^{2}}{2}+\frac{3\kappa }{4},
\label{om-odd}
\end{equation}%
cf. Eq. (\ref{omega}). Note that the special values of $\omega $ necessary
for the existence of the GS and DM exact solutions, as given by Eqs. (\ref%
{omega}) and (\ref{om-odd}), respectively, cannot be equal for $\kappa >0$.
The wave function (\ref{Uodd}) crosses zero at finite values of $x$ under
condition%
\begin{equation}
\lambda ^{2}<(3/2)(1+\kappa ),  \label{0-crossing-2}
\end{equation}%
cf. Eq. (\ref{0-crossing}).

For the exact DM solution, the $P_{u}/P_{v}$ ratio is given by%
\begin{equation}
\left( \frac{P_{u}}{P_{v}}\right) _{\mathrm{DM}}=\frac{\lambda ^{2}}{4}+%
\frac{3(1+\kappa )^{2}}{8\lambda ^{2}},  \label{P/P-DM}
\end{equation}%
cf. Eq. (\ref{P/P}). Accordingly, the $U$ mode is always the dominant one ($%
P_{u}/P_{v}>1$) in the case of%
\begin{equation}
\kappa >2\sqrt{2/3}-1\approx 0.63,  \label{kappa-DM}
\end{equation}%
cf. Eq. (\ref{kappa>}). At smaller values of $\kappa $, there exist exact DM
states with the dominance of the $V$ mode, i.e., with $P_{u}/P_{v}<1$.

\subsection{2D states}

A particular exact solution of 2D stationary equations (\ref{U2D}) and (\ref%
{V2D}) can also be found, for all integer values of vorticity $S$:%
\begin{gather}
U(r)=\left( U_{0}^{\mathrm{(2D)}}+U_{2}^{\mathrm{(2D)}}r^{2}\right)
r^{S}\exp \left( -\frac{r^{2}}{2}\right) ,  \label{U(r)} \\
V(r)=V_{0}^{\mathrm{(2D)}}r^{S}\exp \left( -\frac{r^{2}}{2}\right) ,
\label{V(r)} \\
U_{0}^{\mathrm{(2D)}}=\frac{(S+1)(1+\kappa )-\lambda ^{2}}{2\lambda }V_{0}^{%
\mathrm{(2D)}},  \label{U02D} \\
U_{2}^{\mathrm{(2D)}}=-\frac{1+\kappa }{2\lambda }V_{0}^{\mathrm{(2D)}},
\label{U2-2D} \\
\mu _{\mathrm{2D}}=\frac{1}{2}\left[ \lambda ^{2}+\left( S+1\right) \left(
1-\kappa \right) \right] ,  \label{mu-2D}
\end{gather}%
where $V_{0}^{\mathrm{(2D)}}$ is an arbitrary amplitude. This solution too
is of the codimension-one type, as it exists under the respective constraint:%
\begin{equation}
\omega _{\mathrm{2D}}=\frac{1}{2}\left[ 5+S-\lambda ^{2}+\left( S+1\right)
\kappa \right] .  \label{om2D}
\end{equation}%
It is relevant to mention that the radial wave function (\ref{U(r)}) crosses
zero at finite $r$ under condition $\lambda ^{2}<(S+1)(1+\kappa )$, cf.
similar conditions (\ref{0-crossing}) and (\ref{0-crossing-2}) in the 1D
version of the system.

For the 2D GS solution with $S=0$, the $U$ component dominates, as per Eq. (%
\ref{>1}), at $\kappa >1$, cf. Eqs. (\ref{kappa>}) and (\ref{kappa-DM}). In
the opposite case, the $V$ component is the dominant one in the interval of
\begin{equation}
2-\sqrt{4-(1+\kappa )^{2}}<\lambda ^{2}<2+\sqrt{4-(1+\kappa )^{2}},
\end{equation}%
cf. Eq. (\ref{<lambda<}).

Finally, we note that all the exact solutions given by Eqs. (\ref{U(x)})-(%
\ref{om2D}) can be extended to the case of $\kappa <0$, i.e., for the system
in which both components are subject to the action of trapping potentials.
In particular, codimension-one exact solutions of the half-trapped system,
corresponding to $\kappa =0$, were recently found in Ref. \cite{we}. In the
latter case, all the exact solutions are of the BIC type.

\section{Generic spectra of the bound states}

\subsection{The ground state (GS) of the 1D system}

\subsubsection{The variational approximation (VA)\ }

The codimension-one exact solutions for the trapped modes with propagation
constants (\ref{mu}) and (\ref{mu-odd}), obtained above for the 1D system,
are valid only under constraints (\ref{omega}) or (\ref{om-odd}),
respectively. A possibility to predict generic trapped modes and the
spectrum of their propagation constants is offered by the variational
approximation (VA). Usually, it is applied to solitons and other modes in
nonlinear systems \cite{Progress}, but its original form, known as the
Rayleigh-Ritz approximation, was developed for finding spectra of linear
systems, such as the Schr\"{o}dinger equation in quantum mechanics \cite%
{LL,RR}. In the present context, it is natural to apply it to the prediction
of spectra of bound states in the system of Eqs. (\ref{Ufull}) and (\ref%
{Vfull}) in the linear limit, $\sigma =0$.

In the framework of the VA, 1D wave functions of the GS may be approximated
by the simplest ansatz%
\begin{equation}
\left\{ U_{\mathrm{GS}}^{\mathrm{(VA)}}(x),V_{\mathrm{GS}}^{\mathrm{(VA)}%
}(x)\right\} =\pi ^{-1/4}\left\{ \cos \eta ,\sin \eta \right\} \exp \left( -%
\frac{x^{2}}{2}\right) ,  \label{UV}
\end{equation}%
where $\eta $ is a variational parameter, the Gaussian shape is suggested by
the HO\ trapping potential in Eq. (\ref{Ufull}), and the ansatz is subject
to the normalization condition,%
\begin{equation}
\int_{-\infty }^{+\infty }\left[ \left( U(x)\right) ^{2}+\left( V(x)\right)
^{2}\right] dx=1.  \label{11}
\end{equation}%
Next, one multiplies Eq. (\ref{Ufull}) with $\sigma =0$ by $U(x)$, Eq. (\ref%
{Vfull}) by $V(x)$, integrates each one as $\int_{-\infty }^{+\infty }dx$,
and takes the sum of the two. With regard to normalization condition (\ref%
{11}), this procedure yields an expression for the eigenvalue sought for:%
\begin{eqnarray}
\mu &=&\int_{-\infty }^{+\infty }dx\left\{ U(x)\left[ -\omega U(x)-\frac{1}{2%
}\frac{d^{2}U}{dx^{2}}+\frac{1}{2}x^{2}U(x)\right] \right.  \notag \\
&&\left. +V(x)\left[ -\frac{1}{2}\frac{d^{2}V}{dx^{2}}-\frac{\kappa }{2}%
x^{2}V(x)\right] -2\lambda U(x)V(x)\right\} .  \label{mu-integral}
\end{eqnarray}%
Formally, Eq. (\ref{mu-integral}) gives an exact expression for $\mu $, but
written in terms of unknown wave functions $U(x)$ and $V(x)$. To obtain an
actual result, we insert ansatz (\ref{UV}) and perform the integration,
which yields the following approximate expression:%
\begin{equation}
\mu _{\mathrm{GS}}^{\mathrm{(VA)}}=\left( \frac{1}{2}-\omega \right) \cos
^{2}\eta +\frac{1}{4}\left( 1-\kappa \right) \sin ^{2}\eta -\lambda \sin
\left( 2\eta \right) .  \label{muVA}
\end{equation}%
Now, the VA means minimization of this expression with respect to the free
parameter, $\eta $, i.e., setting%
\begin{equation}
\frac{d}{d\eta }\left( \mu ^{\mathrm{(VA)}}\right) =0.  \label{d/d=0}
\end{equation}%
The substitution of expression (\ref{muVA}) in Eq. (\ref{d/d=0}) produces
the value of $\eta $ sought for:%
\begin{equation}
\tan \left( 2\eta \right) =-\lambda /q_{\mathrm{GS}},  \label{eta}
\end{equation}%
where%
\begin{equation}
q_{\mathrm{GS}}\equiv \frac{1}{2}\left[ \frac{1}{4}\left( 1+\kappa \right)
-\omega \right] .  \label{q}
\end{equation}%
Finally, the substitution of value (\ref{q}) in the VA expression for $\mu $%
, given by Eq. (\ref{muVA}), leads to the following prediction for the
eigenvalue:%
\begin{equation}
\mu _{\mathrm{GS}}^{_{\mathrm{(VA)}}}=\left( \frac{1}{2}-\omega -q_{\mathrm{%
GS}}\right) +\sqrt{q_{\mathrm{GS}}^{2}+\lambda ^{2}}.  \label{final}
\end{equation}

In particular, in the limit of $\kappa \rightarrow \infty $, i.e., in the
limit of the very strong expulsive potential acting upon the $V$ component,
Eq. (\ref{final}) yields%
\begin{equation}
\mu _{\mathrm{GS}}^{_{\mathrm{(VA)}}}\approx \frac{1}{2}-\omega +\frac{%
4\lambda ^{2}}{\kappa },  \label{asympt}
\end{equation}%
which is close to the GS eigenvalue of the $U$ component decoupled from $V$.
Accordingly, in this limit Eqs. (\ref{UV}), (\ref{eta}), and (\ref{q}) show
that the relative amplitude of the $V$ component in the bimodal bound state
decreases as $V/U\approx \eta \approx -4\lambda /\kappa $. This result
explains why the bound state can exist even under the action of the
extremely strong expulsive potential acting upon the $V$ component. On the
other hand, the VA also admits GS solutions dominated by the $V$ component,
i.e., with $P_{v}>P_{u}$, see Eq. (\ref{>1}). The boundary of this situation
corresponds to $\eta =\pi /4$ in ansatz (\ref{UV}), i.e., $q_{\mathrm{GS}}=0$%
, according to Eq. (\ref{eta}). Thus, it follows from Eq. (\ref{q}) that the
VA predicts the domination of the $V$ component at $q_{\mathrm{GS}}<0$.
i.e., at
\begin{equation}
\kappa <4\omega -1.  \label{V domination}
\end{equation}

\subsubsection{Numerical results for the GS}

\begin{figure}[tbp]
\includegraphics[width=12cm]{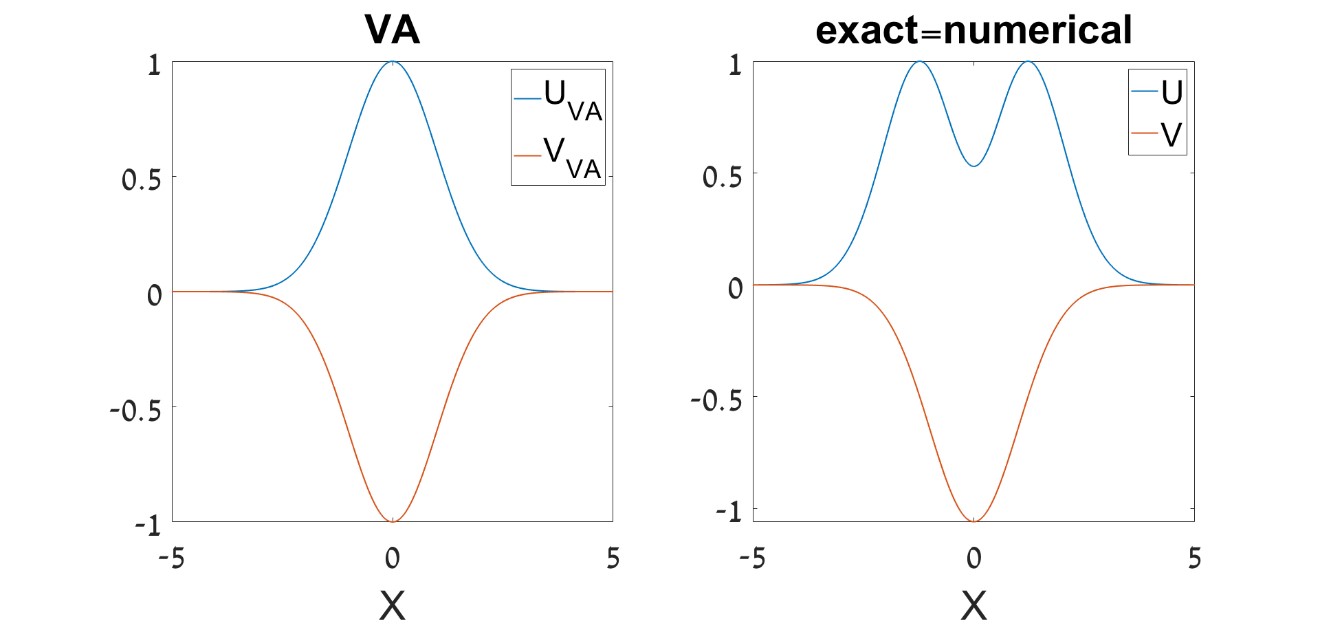}
\caption{The right and left panels display, respectively, exact 1D GS
(ground-state) solution of the linearized ($\protect\sigma =0$) version of
Eqs. (\protect\ref{Ufull}) and (\protect\ref{Vfull}), and its VA-produced
counterpart, at parameter values (\protect\ref{321}). Both the exact
solution and VA yield, in this case, equal eigenvalues, $\protect\mu _{%
\mathrm{GS}}=\protect\mu _{\mathrm{GS}}^{_{\mathrm{(VA)}}}=1.5$. In these
plots, the amplitude of the $V$ component is set to be $V_{0}=-1$. The top
and bottom profiles depict components $U$ and $V$, respectively.}
\label{fig1}
\end{figure}

Comparison of the VA-predicted profile of the GS, produced by Eqs. (\ref{UV}%
), (\ref{11}), (\ref{eta}) and (\ref{q}) at a generic point in the parameter
space,%
\begin{equation}
\kappa =3,\lambda =2,\omega =1,  \label{321}
\end{equation}%
which satisfies condition (\ref{omega}), with the exact profile, as given by
Eqs. (\ref{U(x)})-(\ref{U2}), is displayed in Fig. \ref{fig1}. Unlike the
variational ansatz, the exact analytical solution features a local minimum
at $x=0$, as parameters (\ref{321}) fall in interval (\ref{<<}). In spite of
the discrepancy in the shape, in this case Eqs. (\ref{final}) and (\ref{q})
produce the VA eigenvalue $\mu _{\mathrm{GS}}^{_{\mathrm{(VA)}}}=1.5$, which
\emph{precisely} coincides with the exact eigenvalue given by Eq. (\ref{mu}%
).

\begin{figure}[tbp]
\includegraphics[width=15cm]{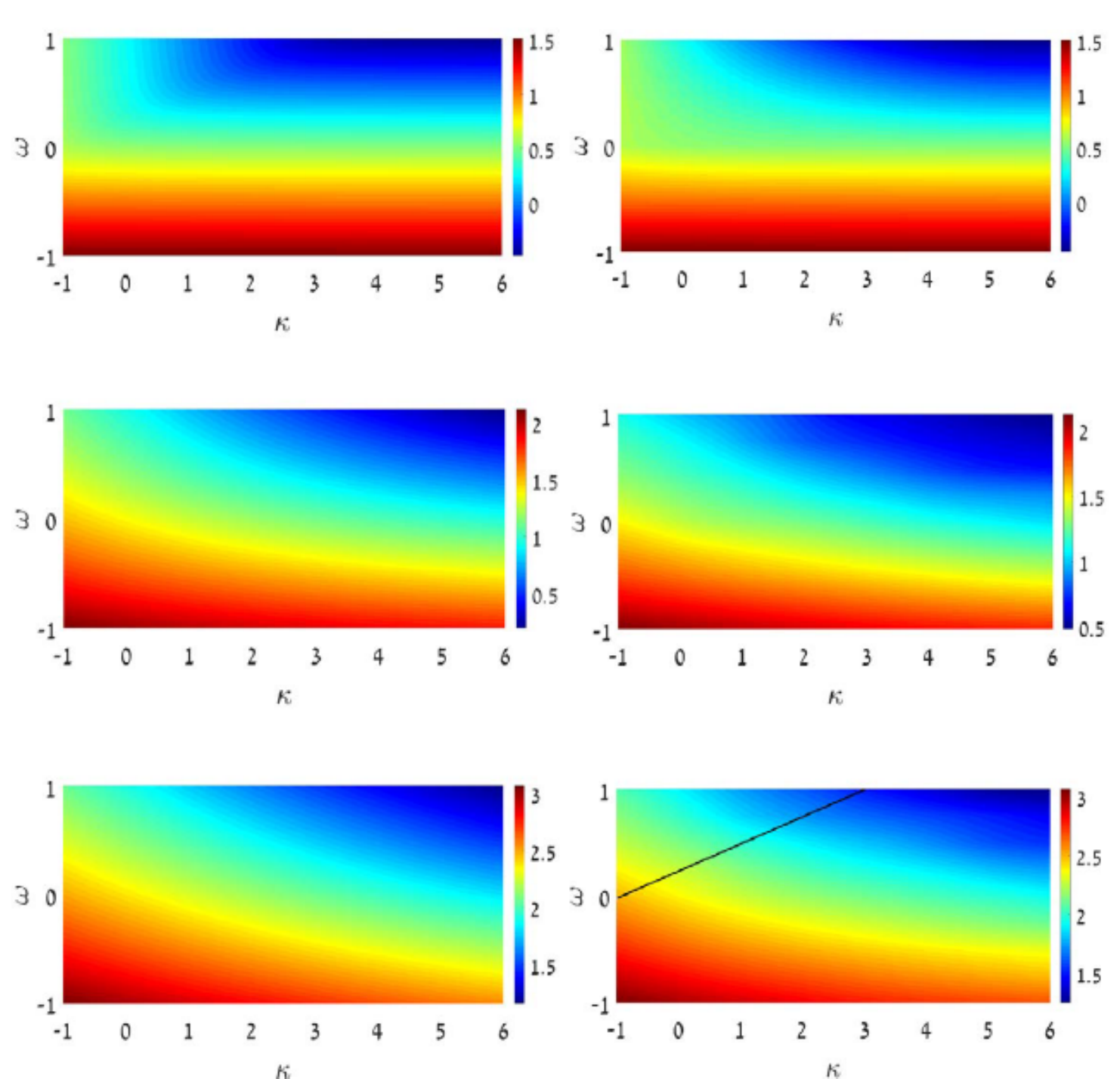}
\caption{Heatmaps (gray-scale maps, in the black-and-white rendition) of the
GS eigenvalue in planes of strength $\protect\kappa $ of the expulsive
potential and mismatch $\protect\omega $ between the components, as produced
by the numerical solution of the linearized equations (\protect\ref{Ufull})
and (\protect\ref{Vfull}), and by the VA (the right and left panels,
respectively). The top, middle, and bottom panels pertain, severally, to
values of the inter-component coupling constant $\protect\lambda =0.1$, $1.0$%
, and $2.0$. The black line in the right bottom panel represents relation (%
\protect\ref{omega}), at which the exact solution (\protect\ref{U(x)})-(%
\protect\ref{mu}) is available.}
\label{fig2}
\end{figure}

The comprehensive results for the eigenvalue of the GS, as predicted by the
VA and produced by the numerical solution of Eqs. (\ref{Ufull}) and (\ref%
{Vfull}) with $\sigma =0$, are summarized, respectively, in the right and
left panels of Fig. \ref{fig2}. It is seen that the trapped GS exists at all
values of the strength of the expulsive potential ($\kappa $). Further, the
VA produces quite accurate results, unless the strength $\lambda $ of the
linear coupling between the components $U$ and $V$, upon which the HO and
anti-HO potentials act, is too small. If $\lambda $ is small, it is not
relevant to adopt the same functional form of both components in the ansatz,
therefore the variational ansatz (\ref{UV}) is inaccurate.

The straight black line in the right bottom panel of Fig. \ref{fig2} is the
locus of points defined by Eq. (\ref{omega}), at which the exact solution is
provided by Eqs. (\ref{U(x)})-(\ref{mu}). Along this line, the eigenvalue of
the numerically found GS solution is precisely equal (up to the numerical
accuracy) to the eigenvalue of the exact solution, as given by Eq. (\ref{mu}%
).

\subsection{The dipole mode (DM) in the 1D system}

\subsubsection{The VA for the DM}

The variational ansatz for the spatially odd DM\ solution is adopted as%
\begin{equation}
\left\{ U_{\mathrm{DM}}^{\mathrm{(VA)}}(x),V_{\mathrm{DM}}^{\mathrm{(VA)}%
}(x)\right\} =\sqrt{2}\pi ^{-1/4}\left\{ \cos \eta ,\sin \eta \right\} x\exp
\left( -\frac{x^{2}}{2}\right) ,  \label{UV-DM}
\end{equation}%
cf. Eq. (\ref{UV}), which is also subject to normalization (\ref{11}).
Substituting this in expression (\ref{mu-integral}) yields%
\begin{equation}
\mu _{\mathrm{DM}}=\left( \frac{3}{2}-\omega \right) \cos ^{2}\eta +\frac{3}{%
4}\left( 1-\kappa \right) \sin ^{2}\eta -\lambda \sin \left( 2\eta \right) ,
\label{muVA-DM}
\end{equation}%
cf. Eq. (\ref{muVA}). Then, the variational equation (\ref{d/d=0}), applied
to expression (\ref{muVA-DM}), produces the result%
\begin{equation}
\tan \left( 2\eta \right) =-\lambda /q_{\mathrm{DM}},  \label{eta-DM}
\end{equation}%
\begin{equation}
q_{\mathrm{DM}}\equiv \frac{1}{2}\left[ \frac{3}{4}\left( 1+\kappa \right)
-\omega \right] ,  \label{q-DM}
\end{equation}%
cf. Eqs. (\ref{eta}) and (\ref{q}). The substitution of this in Eq. (\ref%
{muVA-DM}) leads to the following eigenvalue:%
\begin{equation}
\mu _{\mathrm{DM}}^{\mathrm{(VA)}}=\frac{3}{2}-\omega -q_{\mathrm{DM}}+\sqrt{%
q_{\mathrm{DM}}^{2}+\lambda ^{2}},  \label{final-DM}
\end{equation}%
cf. Eq. (\ref{final}). In particular, in the limit of $\kappa \rightarrow
\infty $, Eq. (\ref{final-DM}) gives%
\begin{equation}
\mu _{\mathrm{DM}}^{\mathrm{(VA)}}\approx \frac{3}{2}-\omega +\frac{4\lambda
^{2}}{3\kappa }.  \label{asympt-DM}
\end{equation}%
Similar to Eq. (\ref{asympt}), the latter result is close to the eigenvalue
of the first excited state of the $U$ component decoupled from $V$.

Also similar to what is considered above for the GS solution, the VA
predicts that the $V$ component is the dominant one in the DM ($P_{v}>P_{u}$%
) at $q_{\mathrm{DM}}<0$, i.e., at $\kappa <(4/3)\omega -1$, cf. Eq. (\ref{V
domination}).

\subsubsection{Numerical results for the DM}

\begin{figure}[tbp]
\includegraphics[width=12cm]{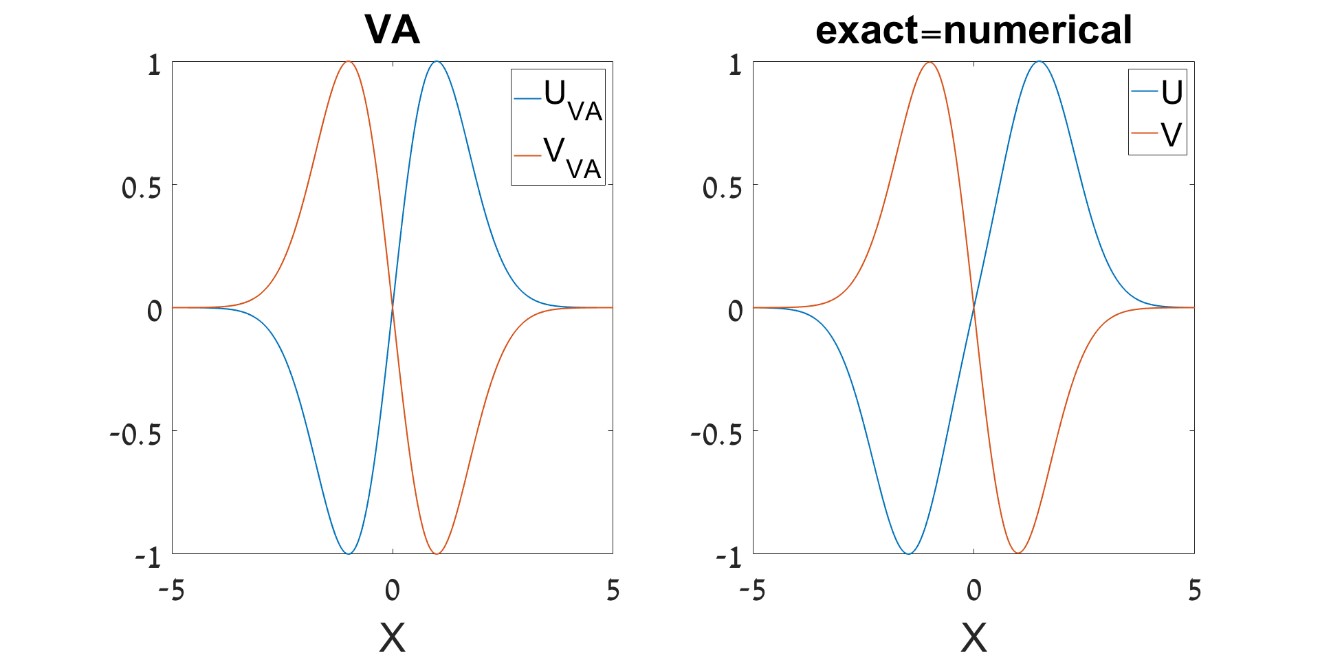}
\caption{The right and left panels display, respectively, the exact 1D DM
(dipole-mode) solution of the linearized ($\protect\sigma =0$) version of
Eqs. (\protect\ref{Ufull}) and (\protect\ref{Vfull}), and its variational
counterpart, for parameters given by Eq. (\protect\ref{1/321}). Both the
exact solution and VA give, in this case, equal eigenvalues, $\protect\mu _{%
\mathrm{DM}}=\protect\mu _{\mathrm{DM}}^{_{\mathrm{(VA)}}}=2.5$. In these
plots, the amplitude of the $V$ component is $V_{0}=-1$. In each panel, the
right and left profiles depict components $U$ and $V$, respectively. }
\label{fig3}
\end{figure}

At characteristic values of parameters,%
\begin{equation}
\kappa =1/3,\lambda =2,\omega =1,  \label{1/321}
\end{equation}%
which satisfy condition (\ref{om-odd}), comparison of the variational
profile of the DM, as produced by Eqs. (\ref{UV-DM}), (\ref{11}), (\ref%
{eta-DM}) and (\ref{q-DM}), with the exact profile, as given by Eqs. (\ref%
{Uodd})-(\ref{U3}) is displayed in Fig. \ref{fig3}. It is worthy to note
that, in comparison with the GS (see Fig. \ref{fig1}), the VA profile for
the DM is closer to its exact counterpart, although the two are not
identical [ratios of amplitudes of the $U$ and $V$ components at parameter
values (\ref{1/321}) are $1$ and $1.004$, as given by the VA and exact
solution, respectively]. Further, in this case Eqs. (\ref{final-DM}) and (%
\ref{q-DM}) produce the VA eigenvalue $\mu _{\mathrm{DM}}^{\mathrm{(VA)}%
}=2.5 $, which precisely coincides with the exact eigenvalue given by Eq. (%
\ref{mu-odd}).

\begin{figure}[tbp]
\includegraphics[width=15cm]{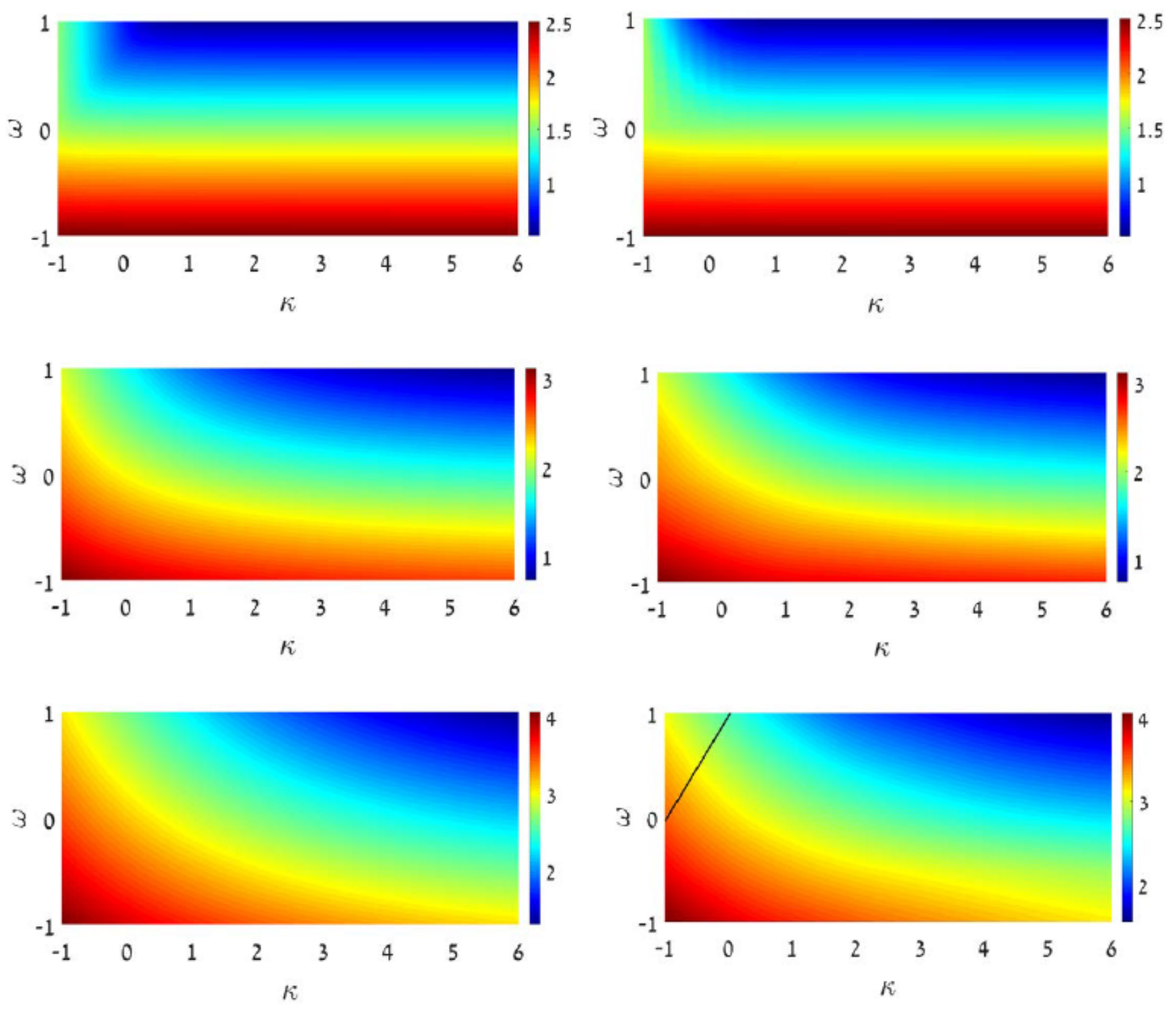}
\caption{Heatmaps (gray-scale maps, in the black-and-white rendition) of the
DM eigenvalue in the $\left( \protect\kappa ,\protect\omega \right) $
planes, as produced by the numerical solution of the linearized equations (%
\protect\ref{Ufull}) and (\protect\ref{Vfull}), and by the VA (right and
left panels, respectively). The top, middle, and bottom panels pertain to
values of the coupling constant $\protect\lambda =0.1$, $1.0$, and $2.0$.
The black line in the right bottom panel represents relation (\protect\ref%
{om-odd}), at which the exact solution is given by Eqs. (\protect\ref{Uodd}%
)-(\protect\ref{mu-odd}).}
\label{fig4}
\end{figure}

Full results for the DM\ eigenvalue, as predicted by the VA, and as produced
by the numerical solution of Eqs. (\ref{Ufull}) and (\ref{Vfull}) with $%
\sigma =0$, are collected in the left and right panels of Fig. \ref{fig4},
respectively. As well as the GS, the DM\ trapped states exist at all values
of the expulsive-potential's strength, $\kappa $. Also similar to what is
observed in Fig. \ref{fig2} for the GS, the VA predicts accurate results,
unless the linear coupling $\lambda $ is very small. The explanation of the
latter point is the same as for the GS, \textit{viz}., ansatz (\ref{UV-DM})
is inaccurate for small $\lambda $.

The straight black line in the right bottom panel of Fig. \ref{fig4} is
defined by Eq. (\ref{om-odd}), as the locus of points at which the exact DM\
solution exists. Along this line, the numerically found GS eigenvalue is
precisely equal to its counterpart given by the exact solution, as per Eq. (%
\ref{mu}).

\subsection{Numerical results for the 2D modes}

\begin{figure}[tbp]
\includegraphics[width=12cm]{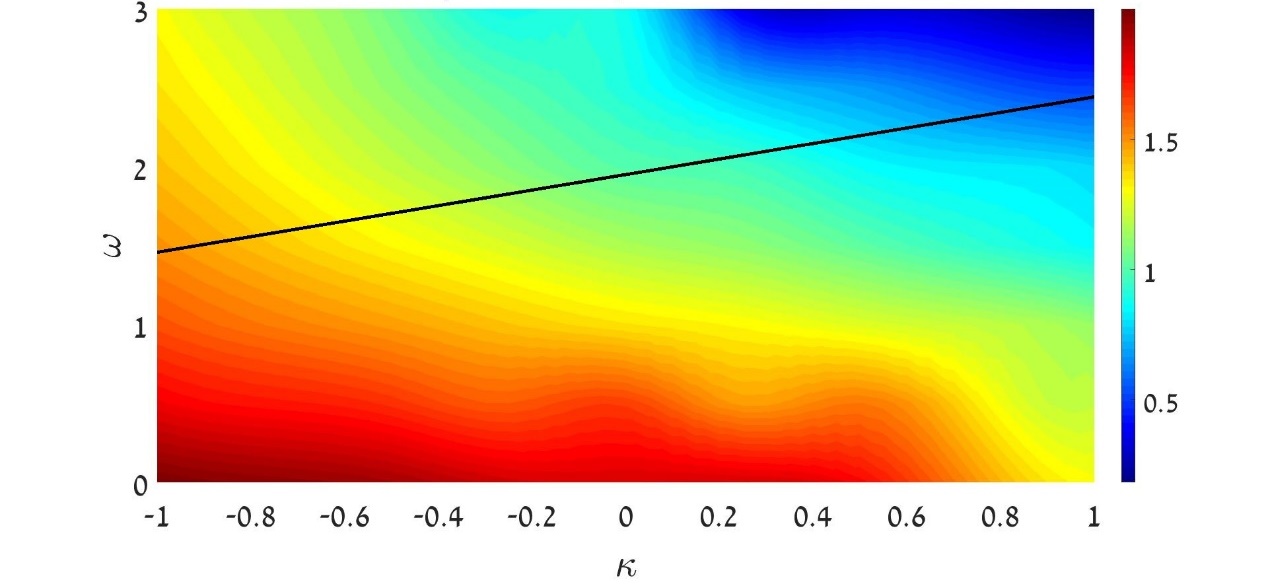}
\caption{The heatmap (gray-scale map, in the black-and-white rendition) of
the eigenvalue of the 2D trapped state with $S=0$ (zero vorticity) in the $%
\left( \protect\kappa ,\protect\omega \right) $ plane at $\protect\lambda =1$%
, as produced by the numerical solution of the linearized equations (\protect
\ref{U2D}) and (\protect\ref{V2D}). The black line represents relation (%
\protect\ref{om2D}), at which the exact 2D solution is given by Eqs. (%
\protect\ref{U(r)})-(\protect\ref{mu-2D}).}
\label{fig5}
\end{figure}

It is commonly known that, due to the separability of the HO potential, the
wave functions trapped in the 2D HO potential can be constructed as products
of 1D eigenstates, and the 2D eigenvalue spectrum can be constructed,
respectively, from its 1D counterpart. However, in the present system the
factorization principle for the eigenstates is broken by linear-coupling
terms. Therefore, it is necessary to compute eigenvalues of the 2D states
numerically. It was thus found that, similar to the 1D case, the solutions
for the bound states (at least, with vorticities $S=0$ and $1$) are produced
by the numerical solution of the linear version of Eqs. (\ref{U2D}) and (\ref%
{V2D}) (with $\sigma =0$) at all values of parameters $\kappa $, $\lambda $,
and $\omega $. A heatmap of the eigenvalues of the 2D states with $S=0$ is
displayed in Fig. \ref{fig5}.

The straight black line crossing the map in Fig. \ref{fig5} is defined by
Eq. (\ref{om2D}). Along this line, the exact solution is given by Eqs. (\ref%
{U(r)})-(\ref{U2-2D}). The respective eigenvalue, given by Eq. (\ref{mu-2D}%
), exactly coincides with its numerically found counterpart.

\subsection{Delocalized states}

The localized states considered above are dominated by the trapped
component, $U$. The same system admits delocalized states, dominated by the
anti-trapped component, $V$. Exact solutions for delocalized states cannot
be found even if the system is taken in the linearized form, with $\sigma =0$%
, but it is possible to construct an asymptotic form of such states at $%
|x|\rightarrow \infty $. The consideration of Eqs. (\ref{Ufull}) and (\ref%
{Vfull}) yields:
\begin{eqnarray}
&&V_{\mathrm{deloc}}(x)\underset{|x|\rightarrow \infty }{\approx }%
V_{0}|x|^{-1/2}\cos \left( \frac{\sqrt{\kappa }}{2}x^{2}+\frac{\mu }{\sqrt{%
\kappa }}\ln \left( |x|\right) \right) ,  \label{V-deloc} \\
&&U_{\mathrm{deloc}}(x)\underset{|x|\rightarrow \infty }{\approx }V_{0}\frac{%
2\lambda }{1+\kappa }|x|^{-5/2}\cos \left( \frac{\sqrt{\kappa }}{2}x^{2}+%
\frac{\mu }{\sqrt{\kappa }}\ln \left( |x|\right) \right) ,  \label{U-deloc}
\end{eqnarray}%
where $V_{0}$ is an arbitrary constant. Obviously, the total power of the $V$%
-component of this solution, defined as per Eq. (\ref{P}), diverges as $\int
dx/|x|$ at $|x|\rightarrow \infty $, therefore such states are categorized
as delocalized ones. The principal difference of the delocalized states from
the trapped ones, in addition to the divergence of the norm, is that they
form a \textit{continuous spectrum}, with the delocalized solutions existing
at all values of $\mu $, as seen from Eqs. (\ref{V-deloc}) and (\ref{U-deloc}%
), while the localized states form the \textit{discrete spectrum}: GS, DM,
etc. Furthermore, the trapped states may be considered as BIC ones, as
discrete values of $\mu $ corresponding to them are embedded into the
continuous spectrum of the delocalized states.

\begin{figure}[tbp]
\subfigure[]{\includegraphics[width=16cm]{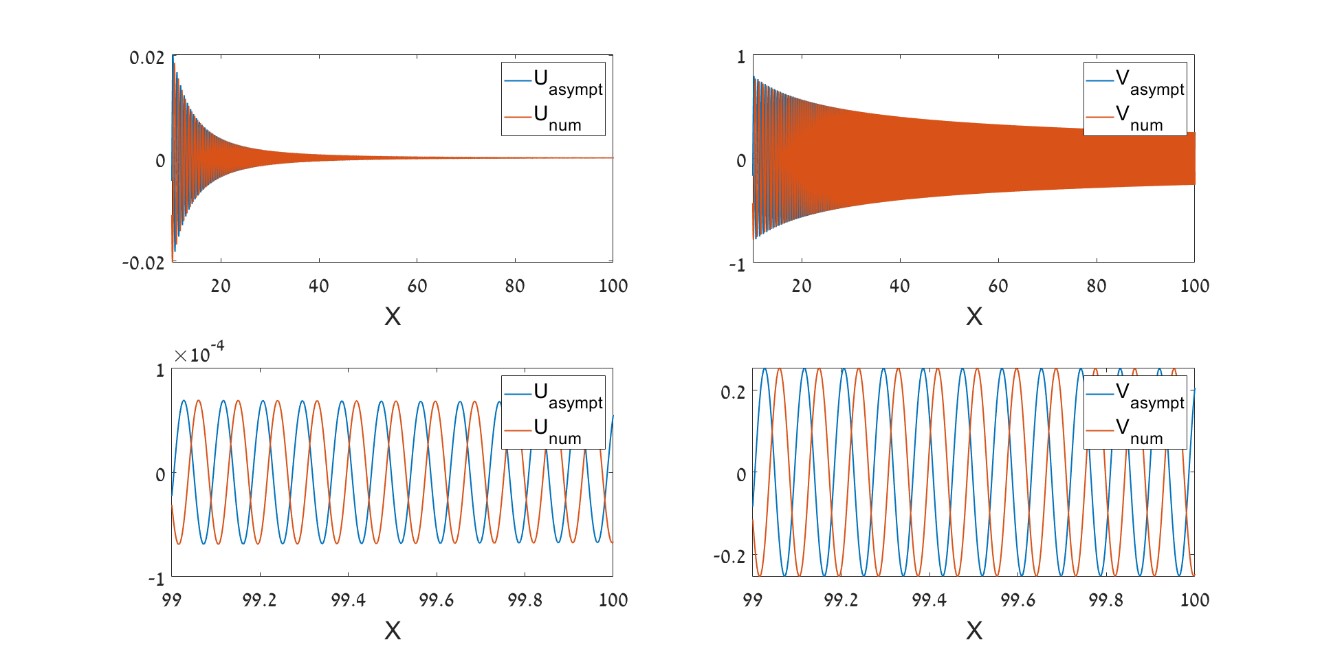}}\newline
\subfigure[]{\includegraphics[width=12cm]{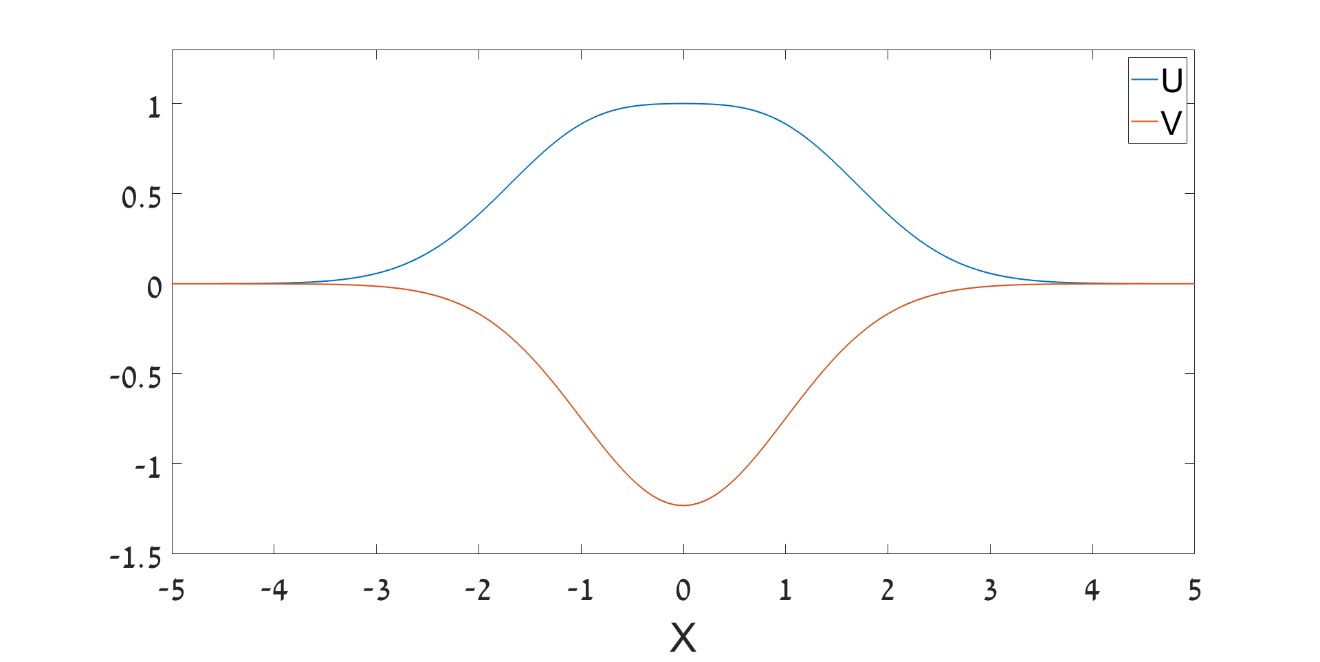}}
\caption{(a) Numerically found shapes of the two components of a delocalized
mode with parameters given by Eq. (\protect\ref{embedded}), and their
counterparts predicted by the asymptotic expressions (\protect\ref{V-deloc})
and (\protect\ref{U-deloc}). (b) The exact GS solution given by Eqs. (%
\protect\ref{U(x)})-(\protect\ref{mu}) for the same parameters, with the top
and bottom profiles depicting $U(x)$ and $V(x)$, respectively.}
\label{fig6}
\end{figure}

A typical example of the delocalized states is displayed in Fig. \ref{fig6}%
(a), along with the analytical asymptotic form, predicted by Eqs. (\ref%
{V-deloc}) and (\ref{U-deloc}), for parameters%
\begin{equation}
\kappa =0.5,\lambda =2,\omega =0.375,\mu =2.125  \label{embedded}
\end{equation}%
These values are chosen because they satisfy Eq. (\ref{omega}), hence the
exact GS solution, given by Eqs. (\ref{U(x)})-(\ref{U2}), is available in
this case, with the corresponding propagation constant $\mu =2.125$ given by
Eq. (\ref{mu}). This GS solution is plotted in Fig. \ref{fig6}(b). Thus, the
existence of the delocalized state with exactly the same value of $\mu $
directly corroborates that the localized states may be considered as ones of
the BIC type. Further, it is seen that the overall shape of the analytical
prediction of the delocalized state is quite accurate, some discrepancy
being represented by a phase shift of the oscillating carrier waves. In
particular, the numerical solution corroborates the fact that the dominant
role is played by the $V$ component.

Similarly, in the 2D case the asymptotic form of solutions to Eqs. (\ref{U2D}%
) and (\ref{V2D}) for delocalized states is%
\begin{eqnarray}
&&V_{\mathrm{deloc}}^{\mathrm{(2D)}}(r)\underset{r\rightarrow \infty }{%
\approx }V_{0}r^{-1}\cos \left( \frac{\sqrt{\kappa }}{2}r^{2}+\frac{\mu }{%
\sqrt{\kappa }}\ln r\right) ,  \label{V2D-deloc} \\
&&U_{\mathrm{deloc}}^{\mathrm{(2D)}}(r)\underset{r\rightarrow \infty }{%
\approx }V_{0}\frac{2\lambda }{1+\kappa }r^{-3}\cos \left( \frac{\sqrt{%
\kappa }}{2}r^{2}+\frac{\mu }{\sqrt{\kappa }}\ln r\right) ,
\label{U2D-deloc}
\end{eqnarray}%
As well as in the 1D case, the 2D delocalized states form a continuous
spectrum, existing at all values of $\mu $, and their total power (norm),%
\begin{equation}
P_{\mathrm{2D}}=2\pi \int_{0}^{\infty }\left[ U^{2}(r)+V^{2}(r)\right] rdr,
\label{N2D}
\end{equation}%
diverges at $r\rightarrow \infty $ as $\int dr/r$.

\section{Bound states and their dynamics in nonlinear systems}

\subsection{Ground states and dipole modes in 1D}

As mentioned above, the results obtained for the linear systems are most
essential for the analysis of the concept of trapped states in the expulsive
potential, maintained by the linear coupling to the component confined by
the trapping potential. Nevertheless, it is also interesting to consider the
role of the nonlinearity in Eqs. (\ref{u}), (\ref{v}) and (\ref{Ufull}), (%
\ref{Vfull}).

\begin{figure}[tbp]
\includegraphics[width=15cm]{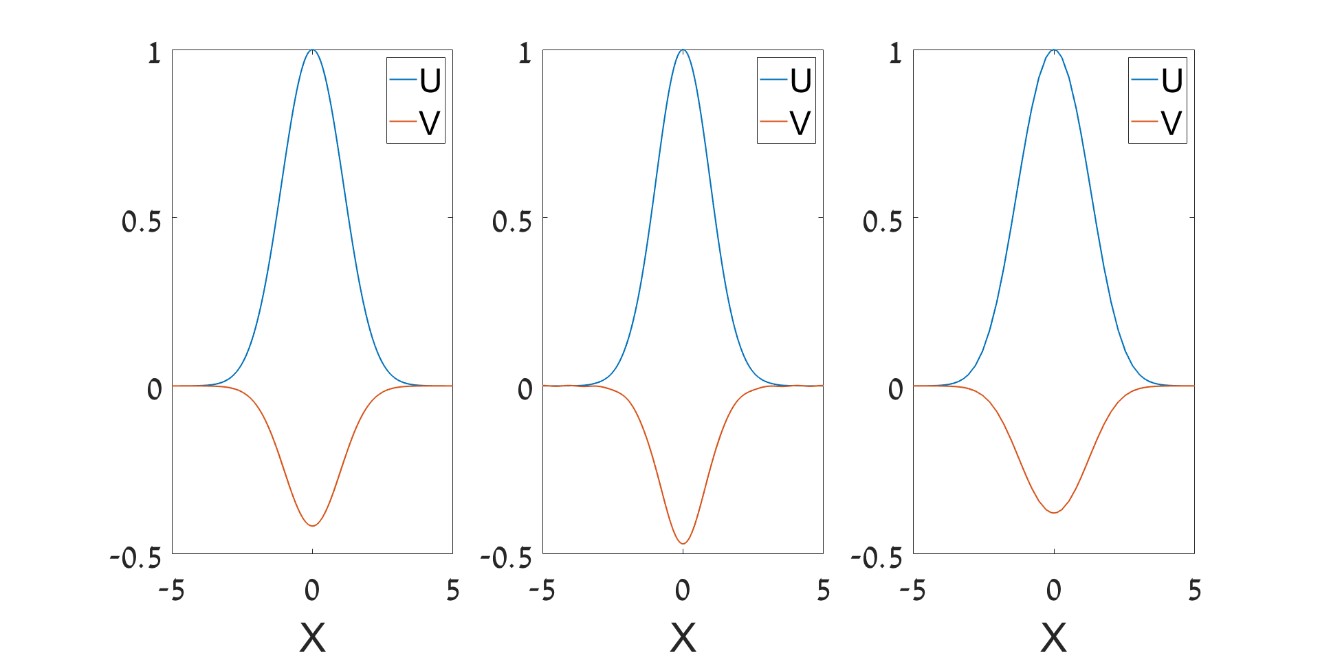}
\caption{The left, central, and right panels show, respectively, the exact
1D GS solution at parameter values (\protect\ref{-10}) and its counterparts
produced by the numerical solution of Eqs. (\protect\ref{Ufull}) and (%
\protect\ref{Vfull}) for $\protect\sigma =+1$ and $-1$. The eigenvalues of
the three solutions are $\protect\mu =12.5$, $11.9735$, and $13.1925$. All
the solutions are produced with the amplitude of the $U$ component $U(x=0)=1$%
. The top and bottom profiles depict components $U(x)$ and $V(x)$,
respectively.}
\label{fig7}
\end{figure}

First, taking into regard the nonlinear terms with $\sigma =+1$ or $-1$
(self-attraction or repulsion) in Eqs. (\ref{Ufull}) and (\ref{Vfull}) leads
to moderate deformation of the 1D GS solution, as shown in Fig. \ref{fig7}
for the set of parameters%
\begin{equation}
\kappa =1,\lambda =5,\omega =-10.  \label{-10}
\end{equation}%
These values satisfy condition (\ref{omega}), hence the solution with $%
\sigma =0$ is given in the exact form by Eqs. (\ref{U(x)})-(\ref{U2}), with
the respective eigenvalue $\mu =12.5$ given by Eq. (\ref{mu}). Naturally,
the self-compression ($\sigma =+1$) and repulsion ($\sigma =-1$) tend to
make the trapped states slightly narrower or broader, respectively. Direct
simulations of the perturbed evolution in the framework of Eqs. (\ref{u})
and (\ref{v}) demonstrate that that the nonlinearly deformed solutions
remain stable (not shown here in detail).

The nonlinear shift of the eigenvalues can be roughly estimated, in the
Thomas-Fermi approximation, by combining Eqs. (\ref{Ufull}) and (\ref{Vfull}%
) at the central point and neglecting the second-derivative terms,
\begin{equation}
\delta \mu \simeq -\sigma \left( U_{0}^{4}+V_{0}^{4}\right) /\left(
U_{0}^{2}+V_{0}^{2}\right) \simeq \mp 0.85  \label{delta}
\end{equation}%
for $\sigma =\pm 1$. This estimate is consistent with the numerically found
shift $\left\vert \delta \mu \right\vert \simeq 0.7$ for the nonlinear
solutions displayed in Fig. \ref{fig7}. Furthermore, the sign of $\delta \mu
$ in Eq. (\ref{delta}) implies that the GS\ family in the nonlinear system
with $\sigma =+1$ satisfies the Vakhitov-Kolokolov (VK) criterion, $d\mu
/dP<0$ [recall $P$ is the total power defined as per Eq. (\ref{P})], which
is the well-known condition necessary for the stability of GS modes in
systems of the nonlinear-Schr\"{o}dinger type with self-attraction \cite%
{VK,Berge,Fibich}. For $\sigma =-1$, Eq. (\ref{delta}) implies $d\mu /dP>0$,
which means that, in the case of the self-repulsion, the GS solutions
satisfy the \textit{anti-VK} criterion, which is a necessary stability
condition for ground states in the case of self-repulsion \cite{HS}.

\begin{figure}[tbp]
\includegraphics[width=12cm]{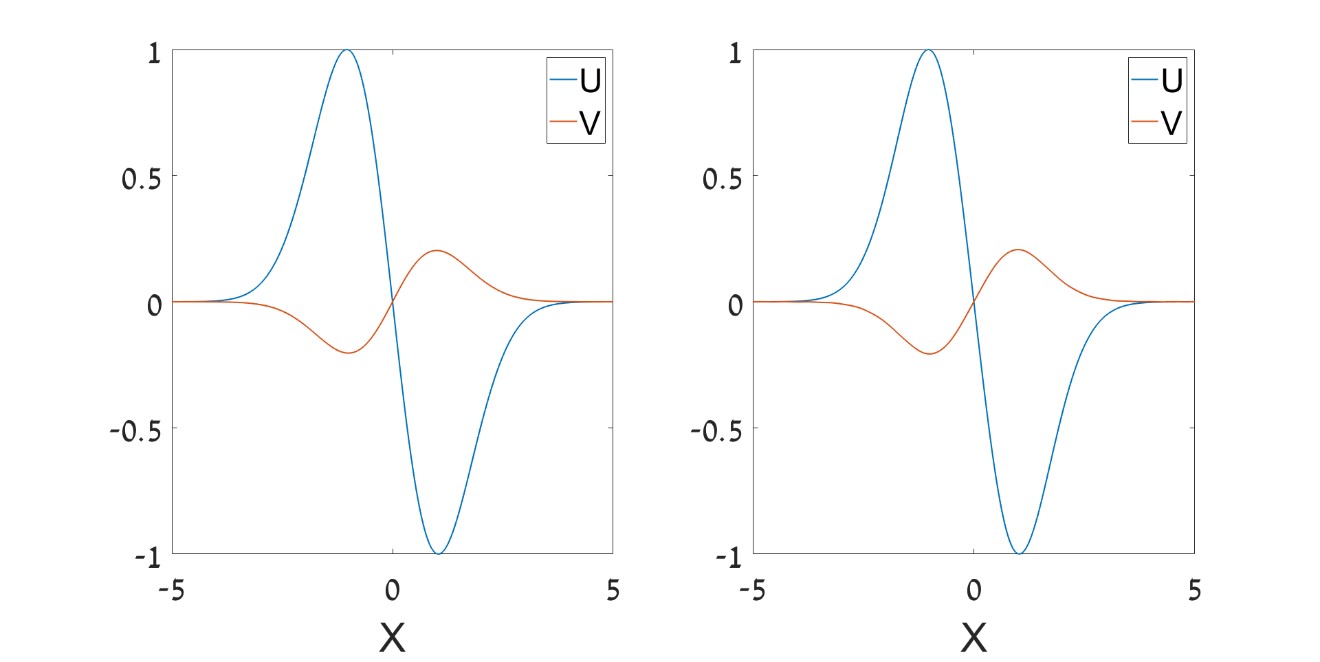}
\caption{The left and right panels show, respectively, the exact 1D DM
solution at parameter values (\protect\ref{-45}) and its counterpart
produced by the numerical solution of Eqs. (\protect\ref{Ufull}) and (%
\protect\ref{Vfull}) for $\protect\sigma =+1$. The eigenvalues of the linear
and nonlinear solutions are, respectively, $\protect\mu =48.875$ and $48.3634
$. The solutions are produced with the amplitude of the $U$ component $%
U_{\max }=1$. The large- and small-amplitude profiles depict components $U$
and $V$, respectively. The solution for $\protect\sigma =-1$ (not shown
here) is very close to these ones.}
\label{fig8}
\end{figure}

The nonlinearity produces a still smaller change of the shape and eigenvalue
of the trapped DM. As an example, Fig. \ref{fig8} displays it for parameters
\begin{equation}
\kappa =2.5,\lambda =10,\omega =-45.375,  \label{-45}
\end{equation}%
which satisfy condition (\ref{om2D}), hence the linear DM\ solution (shown
in the left panel of Fig. \ref{fig8}) is available in the exact form,
according to Eqs. (\ref{Uodd})-(\ref{U3}), with the respective eigenvalue $%
\mu =48.875$ given by Eq. (\ref{mu-odd}). Note that the sign of the
nonlinearity-induced shift of the eigenvalue is the same as in Eq. (\ref%
{delta}).

\begin{figure}[tbp]
\includegraphics[width=12cm]{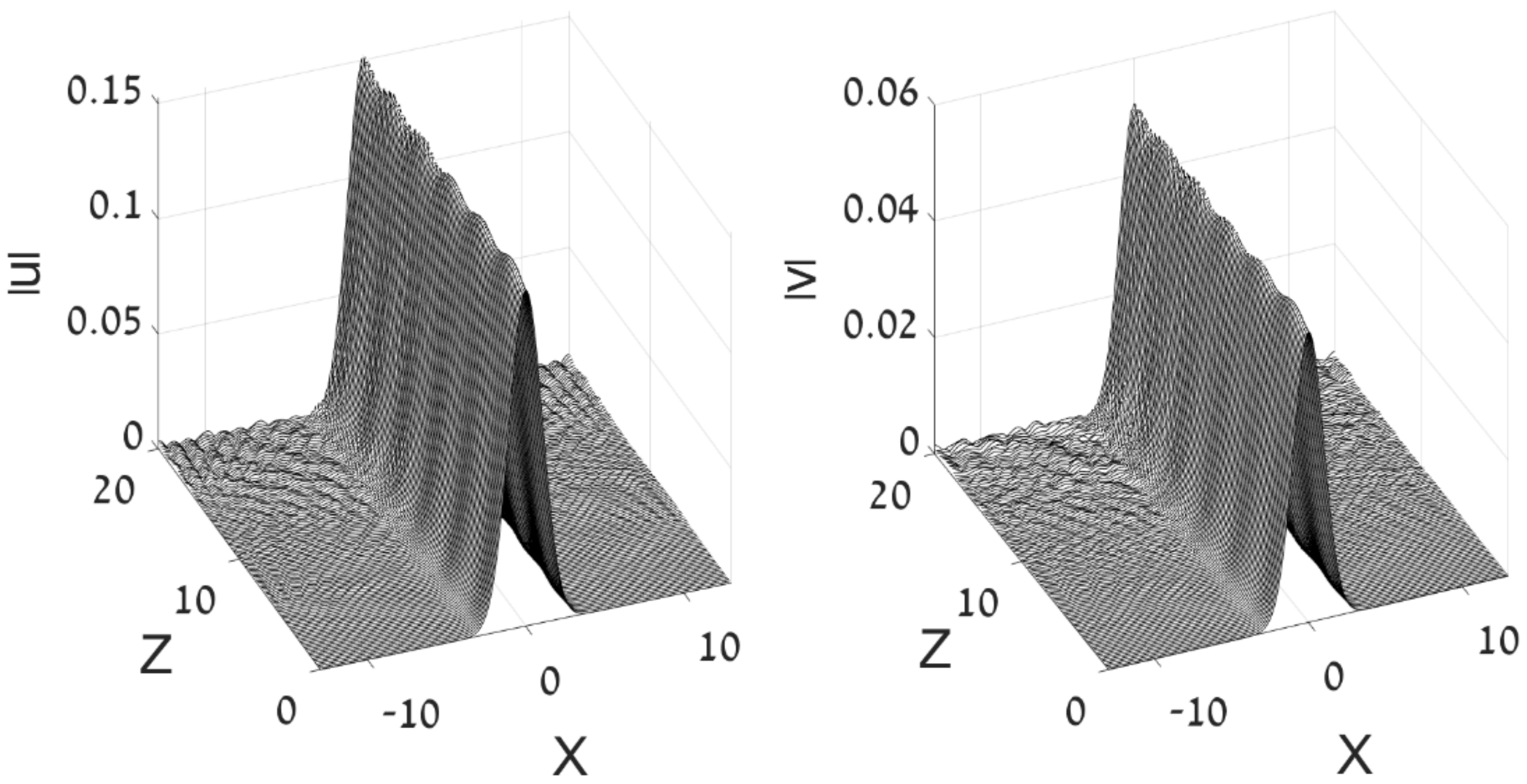}
\caption{Robust oscillations produced by simulations of Eqs. (\protect\ref{u}%
) and (\protect\ref{v}) with parameters (\protect\ref{-15.5}) and $\protect%
\sigma =+1$, and the input given by the linear GS eigenmode (\protect\ref%
{U(x)})-(\protect\ref{U2}) with amplitude $V_{0}=-0.05$ (it corresponds to $%
U_{0}=0.146$).}
\label{fig9}
\end{figure}

\begin{figure}[tbp]
\includegraphics[width=12cm]{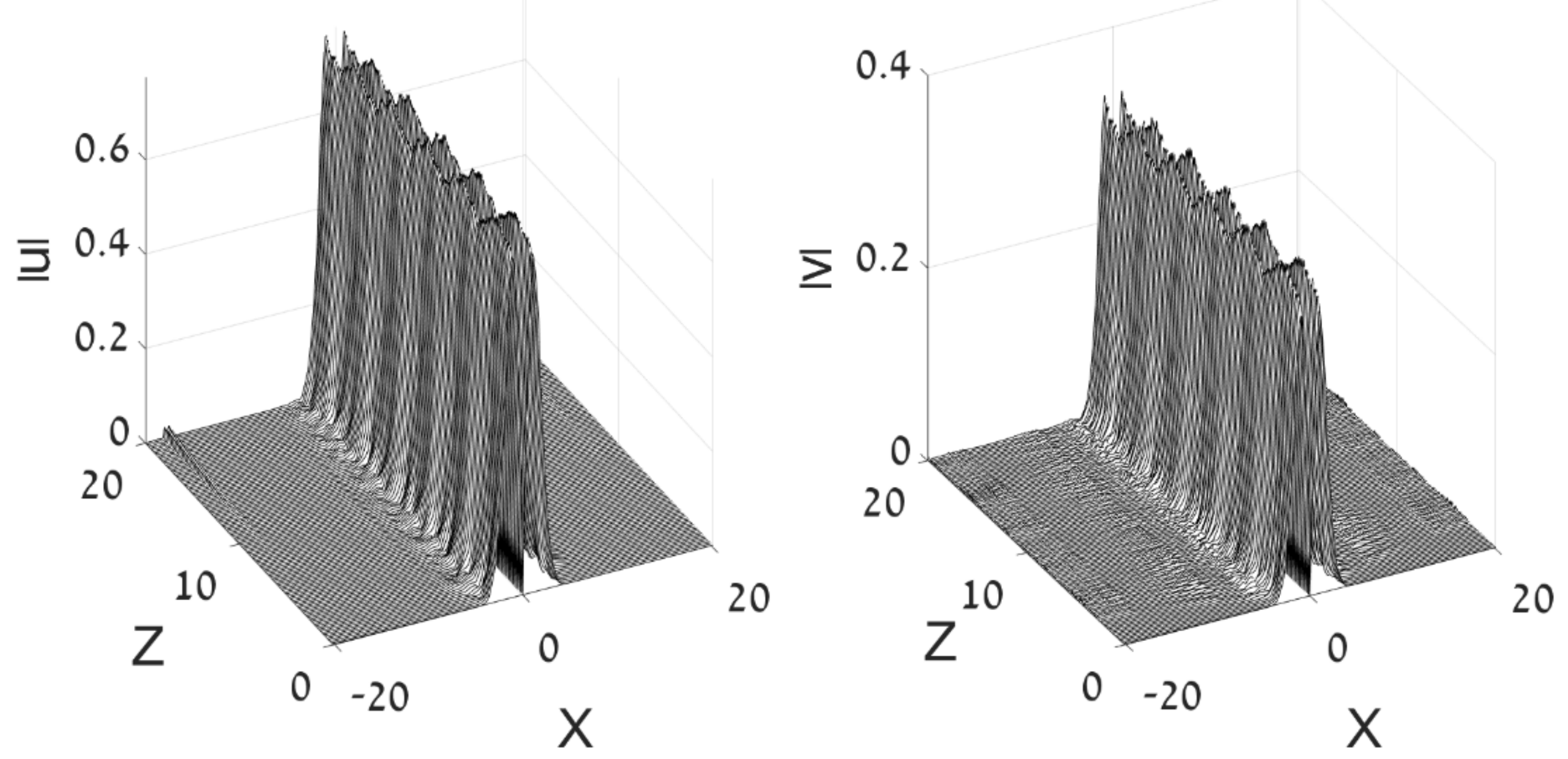}
\caption{The same as in Fig. \protect\ref{fig9}, but for parameters (\protect
\ref{-9.575}) and the input given by the linear DM eigenmode (\protect\ref%
{Uodd})-(\protect\ref{U3}) with amplitude $V_{1}=-0.5$ (it corresponds to
amplitude $U_{\max }=0.716$).}
\label{fig10}
\end{figure}

To explore robustness of bound states in the present setting, it is also
relevant to simulate evolution of an input, taken as an analytical or
numerical eigenstate of the linear system, in the framework of the full
nonlinear system of Eqs. (\ref{u}) and (\ref{v}). The conclusion is that, if
the nonlinearity is moderately strong (in other words, if the input's
amplitude is not too large), the linear eigenstate spontaneously transforms
into a robustly oscillating breather. The respective examples for the GS and
DM are presented, severally, in Fig. \ref{fig9} for
\begin{equation}
\kappa =1,\lambda =6,\omega =-15.5,  \label{-15.5}
\end{equation}%
and in Fig. \ref{fig10} for%
\begin{equation}
\kappa =0.5,\lambda =5,\omega =-9.375.  \label{-9.575}
\end{equation}%

The regular oscillations carry over into chaotic dynamics when the amplitude
exceeds a certain critical value. For instance, the GS input, given by Eqs. (%
\ref{U(x)})-(\ref{U2}) with parameters (\ref{-15.5}) and amplitude $V_{0}$,
evolves into a chaotic state at $\left\vert V_{0}\right\vert >\left(
\left\vert V_{0}\right\vert \right) _{\mathrm{crit}}\simeq 0.75$ for $\sigma
=+1$, and $\left\vert V_{0}\right\vert >\left( \left\vert V_{0}\right\vert
\right) _{\mathrm{crit}}\simeq 1$ for $\sigma =-1$ (these values correspond
to amplitudes $U_{0}\simeq 2.3$ and $U_{0}\simeq 3$, respectively). It is
natural that the chaotic behavior commences later in the case of the
self-defocusing nonlinearity. Similarly, in the case of the DM input given
by Eqs. (\ref{U(x)}-(\ref{U2}) with parameters (\ref{-9.575}), the
chaotization threshold is found at $\left( \left\vert V_{0}\right\vert
\right) _{\mathrm{crit}}\simeq 2$ [which corresponds to $\left( \left\vert
U(x)\right\vert \right) _{\max }\simeq 3$] for both signs of the
nonlinearity, $\sigma =\pm 1$.

An example of the spontaneously established chaos is displayed in Fig. \ref%
{fig11}. It demonstrates that the chaotization effectively destroys the
coupling between the two components; as a result, the $v$ field expands over
the entire integration domain, under the action of the anti-OH potential,
while the $u$ field stays confined by the HO potential acting upon it.
\begin{figure}[tbp]
\includegraphics[width=12cm]{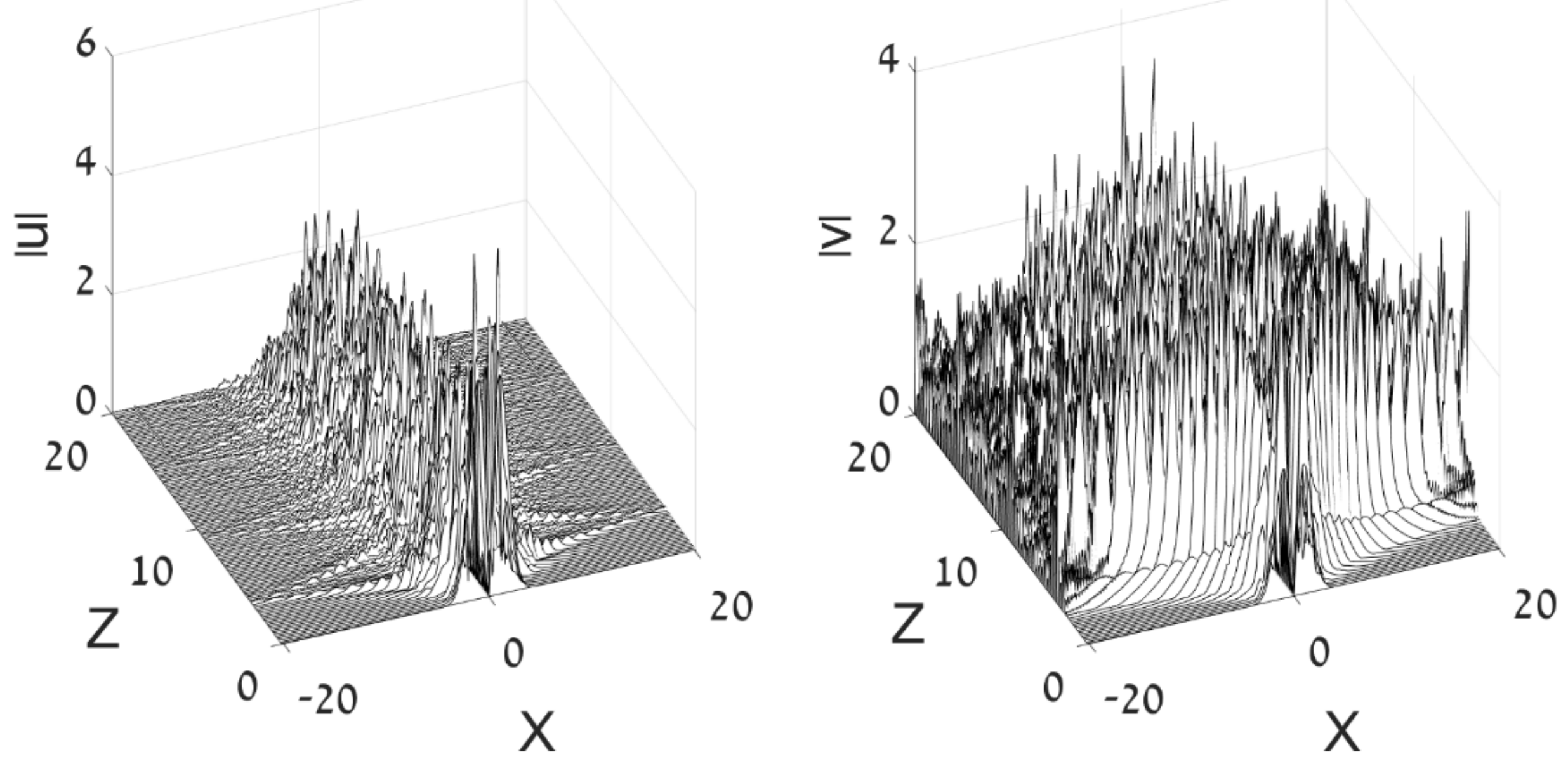}
\caption{The same as in Fig. \protect\ref{fig10}, but for amplitude $%
V_{1}=2.5$ of the DM input.}
\label{fig11}
\end{figure}

\subsection{The ground and vortical states in 2D}

\begin{figure}[tbp]
\includegraphics[width=12cm]{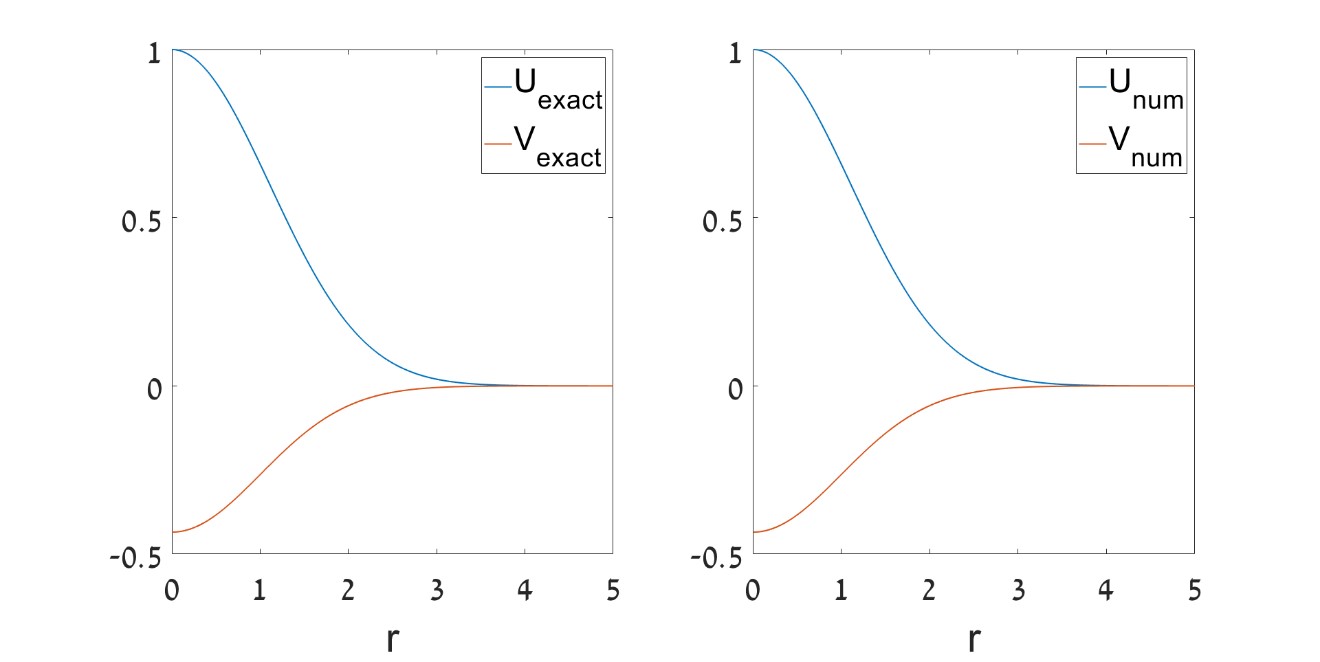}
\caption{The left and right panels display, respectively, the radial profile
of the exact solution (\protect\ref{U(r)})-(\protect\ref{mu-2D}) for the 2D
GS ($S=0$) in the linear system ($\protect\sigma =0$), with parameters (%
\protect\ref{-12}), $\protect\mu =12.5$, and its counterpart produced by the
numerical solution of Eqs. (\protect\ref{U2D}) and (\protect\ref{V2D}) with $%
\protect\sigma =+1$, the respective eigenvalue being $\protect\mu =12.10$.
The amplitude of the $U$ component of both solutions is set to be $U_{0}=1$.
The top and bottom profiles depict components $U(r)$ and $V(r)$,
respectively.}
\label{fig12}
\end{figure}

\begin{figure}[tbp]
\includegraphics[width=12cm]{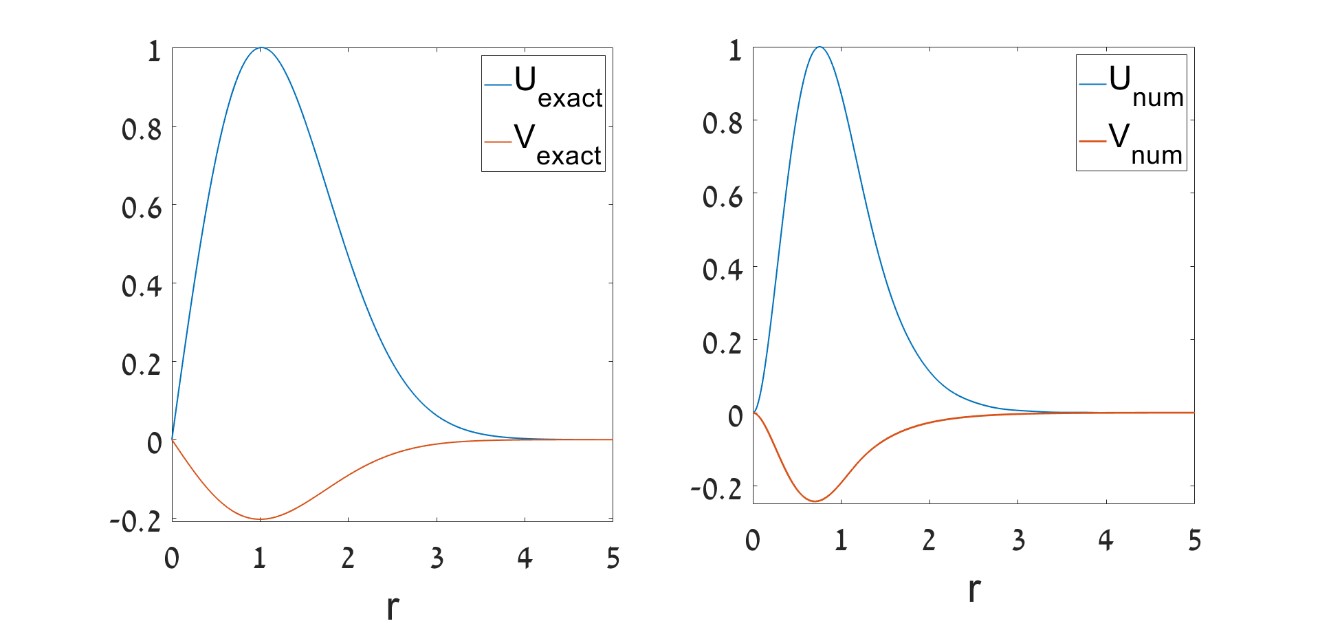}
\caption{The same as in Fig. \protect\ref{fig12}, but for vortex states with
$S=1$ and parameters (\protect\ref{-46.5}). The eigenvalues are $\protect\mu %
=50.5$ and $48.5213$ for the linear and nonlinear solutions, respectively.
The amplitude of the $U$ component of both solutions is set to be $\left(
U(r)\right) _{\max }=1$. The top and bottom profiles depict components $U(r)$
and $V(r)$, respectively.}
\label{fig13}
\end{figure}

As shown in Figs. (\ref{fig12}) and (\ref{fig13}), the moderate nonlinearity
included in 2D stationary equations (\ref{U2D}) and (\ref{V2D}) leads to
weak deformation of the radial profile of the GS ($S=0$) and vortices, and a
very small shift of their propagation constants. In particular, Fig. (\ref%
{fig12}) displays the GS radial profiles for parameters%
\begin{equation}
\kappa =1,\lambda =5,\omega =-12,  \label{-12}
\end{equation}%
which satisfy condition (\ref{om2D}) with $S=0$, thus supplying the exact
solution for $\sigma =0$, as given by Eqs. (\ref{U(r)})-(\ref{mu-2D}).
Note that the difference between the eigenvalues of the nonlinear and linear
solutions in Fig. (\ref{fig12}), $\delta \mu \approx -0.4$, demonstrates
that the 2D GS solutions also satisfy the VK criterion in the case of the
self-focusing nonlinearity. In agreement with this fact, direct simulations
corroborate the stability of these solutions. Similarly, in the case of
self-defocusing, $\sigma =-1$, the nonlinear deformation and eigenvalue
shift are very small too, and the respective GS solutions satisfy the
anti-VK criterion. Their stability was also verified in direct simulations
(not shown here).

Typical examples of the exact linear solution for the vortex mode with $S=1$
and its nonlinear counterpart, produced by the numerical solution of Eqs. (%
\ref{U2D}) and (\ref{V2D}) with $\sigma =+1$, are presented in Fig. \ref%
{fig13}. The parameters are chosen as%
\begin{equation}
\kappa =0.5,\lambda =10,\omega =-46.5,  \label{-46.5}
\end{equation}%
satisfying condition (\ref{om2D}) In this case too, the nonlinearity-induced
deformation of the radial profile and eigenvalue shift, $\delta \mu $, are
relatively small, and the sign of $\delta \mu $ is the same as given by Eq. (%
\ref{delta}).

\begin{figure}[tbp]
\includegraphics[width=16cm]{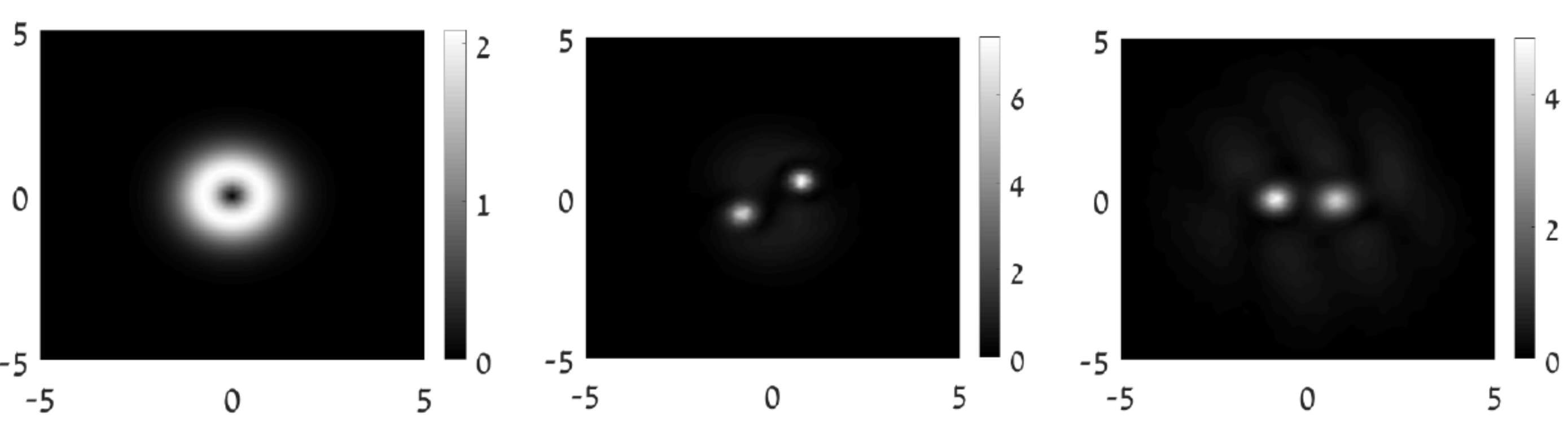}
\caption{Spontaneous splitting of the vortex state with $S=1$, produced by
simulations of Eqs. (\protect\ref{u2D}) and (\protect\ref{v2D}) with
parameters (\protect\ref{-46.5}). The input was provided by the numerical
solution of the stationary equations which is plotted in the right panel of
Fig. \protect\ref{fig13}. The three plots display distributions of density $%
\left\vert u\left( x,y;z\right) \right\vert ^{2}$ at values of the
propagation distance $z=0$ (input), $3.8$, and $6.2$.}
\label{fig14}
\end{figure}

While direct simulations demonstrate that the 2D states with $S=0$ are
completely stable under the action of the nonlinearity of either sign (not
shown here in detail), a well-known problem for localized vortex states in
systems with the self-attractive nonlinearity is their instability against
spontaneous splitting \cite{Quiroga,Alexander,Clark,Raymond} (see also a
review in Ref. \cite{PhysD}). For the present system, we addressed this
issue by means of direct simulations of the evolution of 2D states with $S=1$
in the framework of Eqs. (\ref{u2D}) and (\ref{v2D}). While we did not aim
to explore the entire family of the solutions, the result is that the
vortices are indeed unstable against splitting into a pair of fragments in
the self-focusing system (with $\sigma =+1$), which is a generic outcome of
the instability development known in other systems \cite{PhysD}. A typical
example of the splitting is displayed in Fig. \ref{fig14}. The rotation of
the pair, observed in the figure, provides conservation of the angular
momentum (\ref{M}).

On the other hand, direct simulations demonstrate that vortices with $S=1$
are stable in the self-defocusing system ($\sigma =-1$). An example of the
radial profile of the vortex in this case is plotted (along with its
counterpart provided by the exact solution of the linear system) in Fig. \ref%
{fig15}, for parameters%
\begin{equation}
\kappa =1,\lambda =7,\omega =-20.5,  \label{-20.5}
\end{equation}%
that satisfy Eq. (\ref{om2D}). In this case, the nonlinear deformation and
shift of the propagation constant are quite small too.
\begin{figure}[tbp]
\includegraphics[width=12cm]{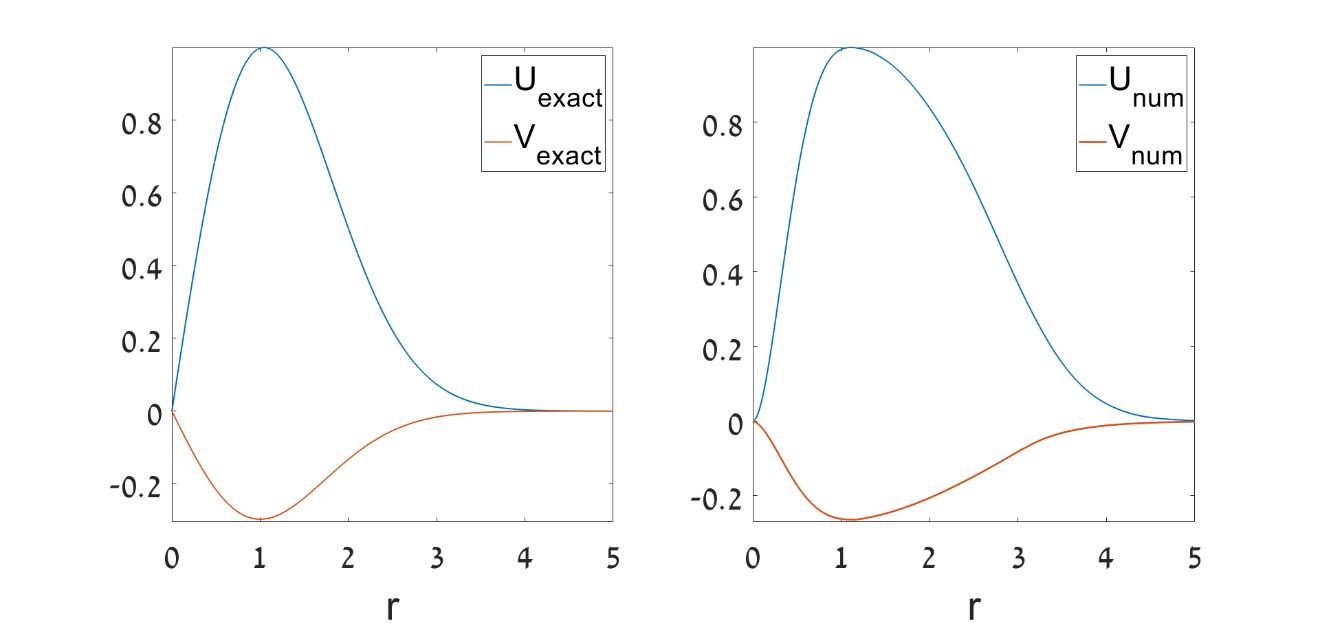}
\caption{The same as in Fig. \protect\ref{fig13}, but for vortex states with
$\protect\sigma =-1$ and parameters (\protect\ref{-20.5}). The eigenvalues
are $\protect\mu =24.5$ and $27.011$ for the linear and nonlinear solutions,
respectively. The amplitude of the $U$ component of both solutions is set to
be $\left( U(r)\right) _{\max }=1$. The top and bottom profiles depict
components $U(r)$ and $V(r)$, respectively.}
\label{fig15}
\end{figure}
Stability of vortices under the action of the self-repulsion is corroborated
by direct simulations, see Fig. \ref{fig16}.
\begin{figure}[tbp]
\includegraphics[width=6cm]{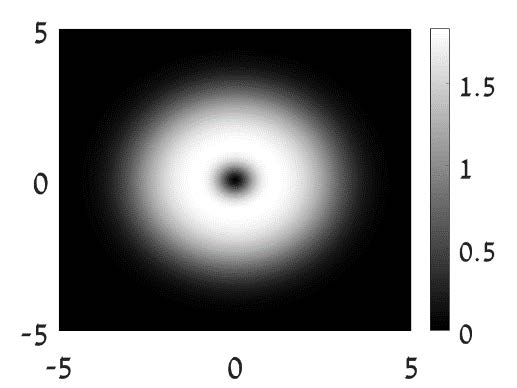} %
\includegraphics[width=6cm]{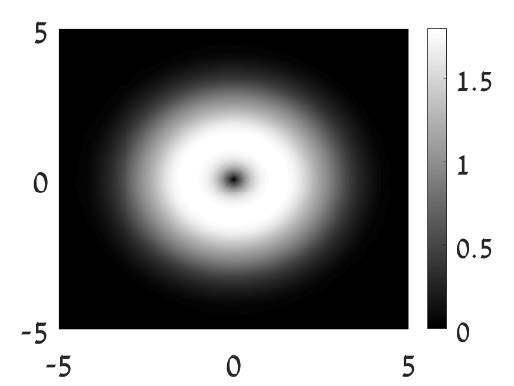}
\caption{Stability of the vortex state with $S=1$, whose stationary profile
is shown in Fig. \protect\ref{fig15}(b), corroborated by simulations of Eqs.
(\protect\ref{u2D}) and (\protect\ref{v2D}) with parameters (\protect\ref%
{-20.5}) and $\protect\sigma =-1$. Left and right plots display,
respectively, distributions of density $\left\vert u\left( x,y;z\right)
\right\vert ^{2}$ at $z=0$ and $10$.}
\label{fig16}
\end{figure}

\section{Conclusion}

The objective of this work is to demonstrate the existence of stable bound
states in the 1D and 2D linearly-coupled two-component systems, with the
trapping HO (harmonic-oscillator) and expulsive anti-HO potentials acting
upon the components. The systems can be realized in optics and BEC. Exact
analytical solutions, which are subject to the codimension-one constraints,
and generic solutions, produced by the VA (variational approximation) and
obtained in the numerical form, demonstrate that the linear systems support
the GS (ground state) and DM (dipole mode, i.e., the first excited state),
as well as 2D vortex states with all integer values of the vorticity, at all
values of the system's parameters. Thus, the linear coupling to the trapped
component makes it possible to maintain robust localized states in the
component subject to the action of the expulsive potential. In most cases,
the trapped component is the dominant one, in terms of the integral power
(norm). Nevertheless, there is a parameter region in which the anti-trapped
component has a larger power. The confined modes supported by this system
may be considered as BIC (bound states in continuum), as they coexist with
delocalized states which form the continuous spectrum. The asymptotic form
of the delocalized solutions is obtained analytically. The inclusion of the
self-attractive or repulsive nonlinearity slightly deforms the 1D states, as
well as the 2D GS, which remain stable solutions. On the other hand, the
self-attraction leads to the splitting instability of vortices, while they
remain stable under the action of self-repulsion.

The work may be extended by considering higher-order eigenstates in 1D, and
performing a more systematic analysis of 2D vortex states. It may also be
interesting to consider coherent motion of self-trapped states in the 1D and
2D systems, cf. Refs. \cite{Belic'2}) and \cite{Busch}.

\section*{Acknowledgment}

This work was supported, in part, by the Israel Science Foundation through
grant No. 1286/17.


\begin{thebibliography}{99}
\bibitem{Pet} C. J. Pethick and H. Smith, \textit{Bose-Einstein Condensation
in Dilute Gases} (Cambridge University Press, Cambridge, 2002).

\bibitem{Pit} L. P. Pitaevskii and S. Stringari, \textit{Bose-Einstein
Condensation} (Oxford University Press, Oxford 2003).

\bibitem{Pan} P. G. Kevrekidis, D. J. Frantzeskakis, and R. Carretero-Gonz%
\'{a}lez, \textit{Emergent Nonlinear Phenomena in Bose-Einstein Condensates:
Theory and Experiment} (Springer: Heidelberg, 2008).

\bibitem{Grossmann} S. Grossmann and M. Holthaus, On Bose-Einstein
condensation in harmonic traps, Phys. Lett. A \textbf{208}, 188-192 (1995).

\bibitem{Castin} Y. Castin and R. Dum, Bose-Einstein condensates in time
dependent traps, Phys. Rev. Lett. \textbf{77}, 5315-5319 (1996).

\bibitem{Fetter} A. L. Fetter and D. L. Feder, Beyond the Thomas-Fermi
approximation for a trapped condensed Bose-Einstein gas, Phys. Rev. A
\textbf{58}, 3185-3194 (1998).

\bibitem{HPu} H. Pu, C. K. Law, J. H. Eberly, and N. P. Bigelow, Coherent
disintegration and stability of vortices in trapped Bose condensates, Phys.
Rev. A \textbf{59}, 1533-1537 (1999).

\bibitem{Schneider} B. I. Schneider and D. L. Feder, Numerical approach to
the ground and excited states of a Bose-Einstein condensed gas confined in a
completely anisotropic trap, Phys. Rev. A \textbf{59}, 2232-2242 (1999).

\bibitem{Sadhan} S. K. Adhikari, Numerical solution of the two-dimensional
Gross-Pitaevskii equation for trapped interacting atoms, Phys. Lett. A
\textbf{265}, 91-96 (2000).

\bibitem{Busch} T. Busch and J. R. Anglin, Motion of dark solitons in
trapped Bose-Einstein condensates, Phys. Rev. Lett. \textbf{84}, 2298-2301
(2000).

\bibitem{Turitsyn} Y. S. Kivshar, T. J. Alexander, and S. K. Turitsyn,
Nonlinear modes of a macroscopic quantum oscillator, Phys. Lett. \textbf{278}%
, 225-230 (2001).

\bibitem{Alexander} T. J. Alexander and L. Berg\'{e}, Ground states and
vortices of matter-wave condensates and optical guided waves, Phys. Rev. E
\textbf{65}, 026611 (2002).

\bibitem{Huang} G. Huang, J. Szeftel and S. Zhu, Dynamics of dark solitons
in quasi-one-dimensional Bose-Einstein condensates, Phys. Rev. A \textbf{65}%
, 053605 (2002).

\bibitem{Nick} N. G. Parker, N. P. Proukakis, M. Leadbeater, and C. S.
Adams, Soliton-sound interactions in quasi-one-dimensional Bose-Einstein
condensates, Phys. Rev. Lett. \textbf{90}, 220401 (2003).

\bibitem{Pelinovsky} D. E. Pelinovsky, D. J. Frantzeskakis, and P. G.
Kevrekidis, Oscillations of dark solitons in trapped Bose-Einstein
condensates, Phys. Rev. E \textbf{72}, 016615 (2005).

\bibitem{Brazhnyi} V. A. Brazhnyi and V. V. Konotop, Stable and unstable
vector dark solitons of coupled nonlinear Schr\"{o}dinger equations:
Application to two-component Bose-Einstein condensates, Phys. Rev. E \textbf{%
72}, 026616 (2005).

\bibitem{Merhasin} M. I. Merhasin, B. A. Malomed, and R. Driben, Transition
to miscibility in a binary Bose-Einstein condensate induced by linear
coupling, J. Physics B \textbf{38}, 877-892 (2005).

\bibitem{Mihalache} D. Mihalache, D. Mazilu, B. A. Malomed, and F. Lederer,
Vortex stability in nearly-two-dimensional Bose-Einstein condensates with
attraction, Phys. Rev. A \textbf{73}, 043615 (2006).

\bibitem{bright-dark} H. E. Nistazakis, D. J. Frantzeskakis, P. G.
Kevrekidis, B. A. Malomed, and R. Carretero-Gonzalez, Bright-dark soliton
complexes in spinor Bose-Einstein condensates, Phys. Rev. A \textbf{77},
033612 (2008).

\bibitem{Nick2} N. G. Parker, N. G. Proukakis, and C. S. Adams, Dark soliton
decay due to trap anharmonicity in atomic Bose-Einstein condensates, Phys.
Rev. A \textbf{81}, 033606 (2010).

\bibitem{Pelin} S. Alama, L. Bronsard, A. Contreras, D. E. Pelinovsky,
Domains walls in the coupled Gross-Pitaevskii equations, Arch. Rat. Mech.
Appl. \textbf{215}, 579-615 (2015).

\bibitem{Viskol} Z. Chen, Y. Li, B. A. Malomed, and L. Salasnich,
Spontaneous symmetry breaking of fundamental states, vortices, and dipoles
in two and one-dimensional linearly coupled traps with cubic
self-attraction, Phys. Rev. A \textbf{96}, 033621 (2017).

\bibitem{Newcastle} T. Bland, N. G. Parker, N. P. Proukakis, and B. A.
Malomed, Probing quasi-integrability of the Gross-Pitaevskii equation in a
harmonic-oscillator potential, J. Phys. B: At. Mol. Opt. Phys. \textbf{51},
205303 (2018).

\bibitem{HS2} H. Sakaguchi and B. A. Malomed, Symmetry breaking in a
two-component system with repulsive interactions and linear coupling, Comm.
Nonlin. Sci. Num. Sim. \textbf{92}, 105496 (2020).

\bibitem{Bizon} P. Bizo\'{n}, F. Ficek, D. E. Pelinovsky, and S. Sobieszek,
Ground state in the energy super-critical Gross-Pitaevskii equation with a
harmonic potential, Nonlinear Anal. \textbf{210}, 112358 (2021).

\bibitem{Evnin} A. Biasi, O. Evnin, and B. A. Malomed, Fermi-Pasta-Ulam
phenomena and persistent breathers in the harmonic trap, Phys. Rev. E
\textbf{104}, 034210 (2021).

\bibitem{PLA} B. A. Malomed, New findings for the old problem: Exact
solutions for domain walls in coupled real Ginzburg-Landau equations, Phys.
Lett. A \textbf{432}, 127802 (2021).

\bibitem{Agrawal} S. Raghavan and G. P. Agrawal, Spatiotemporal solitons in
inhomogeneous nonlinear media, Opt. Commun. \textbf{180}, 377-382 (2000).

\bibitem{Zezyu} D. A. Zezyulin, G. L. Alfimov, and V. V. Konotop, Nonlinear
modes in a complex parabolic potential, Phys. Rev. A \textbf{81}, 013606
(2010).

\bibitem{Stathis} E. G. Charalampidis, P. G. Kevrekidis, D. J.
Frantzeskakis, and B. A. Malomed, Dark-bright solitons in coupled NLS
equations with unequal dispersion coefficients, Phys. Rev. E \textbf{91},
012924 (2015).

\bibitem{Belic'} Y. Zhang, X. Liu, M. R. Beli\'{c}, W. Zhong, Y. Zhang, and
M. Xiao, Propagation dynamics of a light beam in a fractional Schr\"{o}%
dinger equation, Phys. Rev. Lett. \textbf{115}, 180403 (2015).

\bibitem{Thaw} T. Mayteevarunyoo, B. A. Malomed, and D. V. Skryabin, One-
and two-dimensional modes in the complex Ginzburg-Landau equation with a
trapping potential, Opt. Exp. \textbf{26}, 8849-8865 (2018).

\bibitem{polaritons} R. Balili, V. Hartwell, D. Snoke, L. Pfeiffer, and K.
West, Bose-Einstein condensation of microcavity polaritons in a trap,
Science \textbf{316}, 1007-1010 (2007).

\bibitem{Josephson network} M. Leib, F. Deppe, A. Marx, R. Gross, and M. J.
Hartmann, Networks of nonlinear superconducting transmission line
resonators, New J. Phys. \textbf{14}, 075024 (2012).

\bibitem{KA} Y. S. Kivshar and G. P. Agrawal, \textit{Optical Solitons: From
Fibers to Photonic Crystals }(Academic Press: San Diego, 2003).

\bibitem{Peyrard} T. Dauxois, M. Peyrard, \textit{Physics of Solitons}
(Cambridge University Press: Cambridge, 2006).

\bibitem{Carr} L. D. Carr and Y. Castin, Dynamics of a matter-wave bright
soliton in an expulsive potential, Phys. Rev. A \textbf{66}, 063602 (2002).

\bibitem{Salasnich} L. Salasnich, Dynamics of a Bose-Einstein-condensate
bright soliton in an expulsive potential, Phys. Rev. A \textbf{70}, 053617
(2004).

\bibitem{Liu} Z. X. Liang, Z. D. Zhang, and W. M. Liu, Dynamics of a bright
soliton in Bose-Einstein condensates with time-dependent atomic scattering
length in an expulsive parabolic potential, Phys. Rev. Lett. \textbf{94},
050402 (2005).

\bibitem{Cardoso} W. B. Cardoso, A. T. Avelar, and D. Bazeia, Bright and
dark solitons in a periodically attractive and expulsive potential with
nonlinearities modulated in space and time, Nonlin. Analysis Real World
Appl. \textbf{11}, 4269-4274 (2010).

\bibitem{Lakshmanan} S. Rajendran, P. Muruganandam, M. Lakshmanan, Bright
and dark solitons in a quasi-1D Bose-Einstein condensates modelled by 1D
Gross-Pitaevskii equation with time-dependent parameters, Physica D 239,
366-386 (2010).

\bibitem{XuanXuo} N.H. Xuan, M. Zuo, Matter-wave solitons in two-component
Bose-Einstein condensates with tunable interactions and time varying
potential, Comm. Theor. Phys. \textbf{56}, 1035-1040 (2011).

\bibitem{Lakshmanan2} S. Rajendran, M. Lakshmanan, P. Muruganandam, Matter
wave switching in Bose-Einstein condensates via intensity redistribution
soliton interactions, J. Math. Phys. 52, 023515 (2011).

\bibitem{Boshier} A. del Campo, M.G. Boshier, Shortcuts to adiabaticity in a
time-dependent box, Sci. Rep. \textbf{2}, 648 (2012).

\bibitem{Radha} R. Radha, P. S. Vinayagam, J. B. Sudharsan, and B. A.
Malomed, Persistent bright solitons in sign-indefinite coupled nonlinear Schr%
\"{o}dinger equations with a time-dependent harmonic trap, Comm. Nonlin.
Science Num. Sim. \textbf{31}, 30-39 (2016).

\bibitem{KartKon} Y. V. Kartashov, V. V. Konotop, Stable nonlinear modes
sustained by gauge fields, Phys. Rev. Lett. \textbf{125}, 054101 (2020).

\bibitem{anti-WG1} B. V. Gisin and A. A. Hardy, Stationary solutions of
plane nonlinear-optical antiwaveguides, Opt. Quant. Elect. \textbf{27},
565-575 (1995).

\bibitem{anti-WG2} B. V. Gisin, A. Kaplan, and B. A. Malomed, Spontaneous
symmetry breaking and switching in planar nonlinear optical antiwaveguides,
Phys. Rev. E \textbf{62}, 2804-2809 (2000).

\bibitem{anti-WG3} D. Bortman-Arbiv, A. D. Wilson-Gordon, and H. Friedmann,
Strong parametric amplification by spatial soliton-induced cloning of
transverse beam profiles in an all-optical antiwaveguide, Phys. Rev. A
\textbf{63}, 031801(R) (2001).

\bibitem{anti-WG4} O. N. Verma and T. N. Dey, Steering, splitting, and
cloning of an optical beam in a coherently driven Raman gain system, Phys.
Rev. A \textbf{9}1, 013820 (2015).

\bibitem{Kaplan} A. Kaplan, B. V. Gisin, and B. A. Malomed, Stable
propagation and all-optical switching in planar waveguide-antiwaveguide
periodic structures. J. Opt. Soc. Am. B \textbf{19}, 522-527 (2002).

\bibitem{HS} H. Sakaguchi and B. A. Malomed, Dynamics of positive- and
negative-mass solitons in optical lattices and inverted traps, J. Phys. B
\textbf{37}, 1443-1459 (2004).

\bibitem{Ketterle} Y. Shin, G.-B. Jo, M. Saba, T. A. Pasquini, W. Ketterle,
and D. E. Pritchard, Optical weak link between two spatially separated
Bose-Einstein condensates, Phys. Rev. Lett. \textbf{95}, 170402 (2005).

\bibitem{two-color} S. J. Park, J. Noh, and J. Mun, Cold atomic beam from a
two-dimensional magneto-optical trap with two-color pushing laser beams,
Opt. Commun. \textbf{285}, 3950-3954 (2012).

\bibitem{polymers} H. Ma, A. K. Y. Jen, and L. R. Dalton, Polymer-based
optical waveguides: Materials, processing, and devices, Advanced Materials
\textbf{14}, 1339-1365 (2002).

\bibitem{we} N. Hacker and B. A. Malomed, Nonlinear dynamics of wave packets
in tunnel-coupled harmonic-oscillator traps, Symmetry \textbf{13}, 372
(2021).

\bibitem{BIC} F. H. Stillinger and D. R. Herrick, Bound states in continuum,
Phys. Rev. A \textbf{11}, 446-454 (1975).

\bibitem{BIC2} A. Kodigala, T. Lepetit, Q. Gu, B. Bahari, Y. Fainman, and B.
Kante, Lasing action from photonic bound states in continuum, Nature \textbf{%
54}1, 196-199 (2017).

\bibitem{SOC} Y. V. Kartashov, V. V. Konotop, and L. Torner, Bound states in
the continuum in spin-orbit-coupled atomic systems, Phys. Rev. A \textbf{96}%
, 033619 (2017).

\bibitem{BIC3} Y. Guo, M. Xiao, S. Fan, Topologically protected complete
polarization conversion, Phys. Rev. Lett. \textbf{119}, 167401 (2017).

\bibitem{BIC6} B. Midya, and V. V. Konotop, Coherent-perfect-absorber and
laser for bound states in a continuum, Opt. Lett. \textbf{43}, 607-610
(2018).

\bibitem{BIC4} L. Xu, K. Z. Kamali, L. Huang, M. Rahmani, A. Smirnov, R.
Camacho-Morales, Y. X. Ma, G. Q. Zhang, M. Woolley, D. Neshev, and A. E.
Miroshnichenko, Dynamic nonlinear image tuning through magnetic dipole
quasi-BIC ultrathin resonators, Advanced Science \textbf{6}, 1802119 (2019).

\bibitem{BIC5} S. Romano, G. Zito, S. N. L. Yepez, S. Cabrini, E. Penzo, G.
Coppola, I. Rendina, and V. Mocella, Tuning the exponential sensitivity of a
bound-state-in-continuum optical sensor, Opt. Exp. \textbf{27}, 18776-18786
(2019).

\bibitem{embedded} A. R. Champneys, B. A. Malomed, J. Yang, and D. J. Kaup,
\textquotedblleft Embedded solitons": solitary waves in resonance with the
linear spectrum, Physica D \textbf{152-153}, 340-354 (2001).

\bibitem{PhysD} B. A. Malomed, (INVITED)\ Vortex solitons: Old results and
new perspectives, Physica D \textbf{399}, 108-137 (2019).

\bibitem{Avila} A. Avila, Global theory of one-frequency Schr\"{o}dinger
operators, Acta Mathematica 215, 1-54 (2015).

\bibitem{Progress} B. A. Malomed, Variational methods in nonlinear fiber
optics and related fields, Progr. Optics \textbf{43}, 71-193 (2002).

\bibitem{LL} L. D. Landau and E. M. Lifshitz, \textit{Quantum Mechanics}
(Nauka Publishers, Moscow, 1974).

\bibitem{RR} R. N. Hill. Rates of convergence and error estimation formulas
for the Rayleigh-Ritz variational method, J. Chem. Phys. \textbf{83},
1173-1196 (1985).

\bibitem{VK} N. G. Vakhitov and A. A. Kolokolov, Stationary solutions of the
wave equation in a medium with nonlinearity saturation, Radiophys. Quant.
Electron. \textbf{16}, 783-789 (1973).

\bibitem{Berge} L. Berg\'{e}, Wave collapse in physics: principles and
applications to light and plasma waves, Phys. Rep. \textbf{303}, 259-370
(1998).

\bibitem{Fibich} G. Fibich, \textit{The Nonlinear Schr\"{o}dinger Equation:
Singular Solutions and Optical Collapse} (Springer: Heidelberg, 2015).

\bibitem{anti} H. Sakaguchi and B. A. Malomed, Solitons in combined linear
and nonlinear lattice potentials, Phys. Rev. A \textbf{81}, 013624 (2010).

\bibitem{Quiroga} M. Quiroga-Teixeiro and H. Michinel, Stable azimuthal
stationary state in quintic nonlinear optical media, J. Opt. Soc. Amer. B
\textbf{14}, 2004-2009 (1997).

\bibitem{Clark} L. D. Carr and C. W. Clark, Vortices in attractive
Bose--Einstein condensates in two dimensions, Phys. Rev. Lett. \textbf{97}
010403 (2006).

\bibitem{Raymond} Y. Li, Z. Chen, Z. Luo, C. Huang, H. Tan, W. Pang, and B.
A. Malomed, Two-dimensional vortex quantum droplets, Phys. Rev. A \textbf{98}%
, 063602 (2018).

\bibitem{Belic'2} M. R. Beli\'{c}, A. Stepken, and F. Kaiser, Spatial
screening solitons as particles, Phys. Rev. Lett. \textbf{84}, 83-86 (2000).
\end{thebibliography}
\end{document}